\documentclass[universe,review,accept,pdftex,oneauthor]{Definitions/mdpi}
\usepackage{amssymb}
\usepackage{Definitions/aas_macros}
\firstpage{1} 
\pubvolume{10}
\issuenum{6}
\articlenumber{260}
\pubyear{2024}
\copyrightyear{2024}
\externaleditor{Academic Editor: Binbin Zhang}
\datereceived{24 April 2024} 
\daterevised{21 May 2024} 
\dateaccepted{29 May 2024} 
\datepublished{12 June 2024} 
\hreflink{https://\linebreak doi.org/10.3390/universe10060260} 

\Title{{A Short} 
 History of the First 50 Years: From the GRB Prompt Emission and Afterglow Discoveries to the Multimessenger Era}

\TitleCitation{A Short History of the First 50 Years: From the GRB Prompt Emission and Afterglow Discoveries to the Multimessenger Era}

\Author{{Filippo} 
 Frontera $^{1,2}$\orcidA{}}

\AuthorNames{Filippo Frontera}

\AuthorCitation{Frontera, F.}

\address{%
$^{1}$ \quad {Department of} 
 Physics and Earth Science, University of Ferrara, {44122 Ferrara}
, Italy; frontera@fe.infn.it\\
$^{2}$ \quad {National Institute of Astrophysics (INAF), Astrophysics and Space Science Observatory (OAS),} 
 {Via~Gobetti 101,} 
 \mbox{40129 Bologna, Italy}}

\def \vela{{\em Vela}\/}
\def \venera{{\em Venera}\/}
\def \konus{{\em Konus}}
\def \hete{{\em HETE-2}\/}
\def \rosat{{\em ROSAT}\/}
\def \sax{{\em BeppoSAX}\/}
\def \grbm{{\bf GRBM}\/}
\def \wfc{{ WFC}\/}
\def \wfcs{{ WFCs}\/}
\def \bat{{BAT}\/}
\def \gbm{{ GBM}\/}
\def \swift{{\em Swift}\/}
\def \agile{{\em AGILE}\/}

\def \fermi{{\em Fermi}\/}
\def \cgro{{\em Compton Gamma Ray Observatory}\/}
\def \batse{{BATSE}\/}
\def \integral{{\em INTEGRAL}\/}
\def \nustar{{\em NuSTAR}\/}
\def \magic{{\em MAGIC}\/}
\def \hess{{\em H.E.S.S.}\/}
\def \hst{{\em Hubble Space Telescope}\/}
\def \theseus{{\em THESEUS}\/}

\abstract{More than fifty years have elapsed from the first discovery of gamma-ray bursts (GRBs) with American {\vela\ }
 satellites, and more than twenty-five years from the discovery with the {\sax\ }satellite of the first X-ray afterglow of a GRB. Thanks to the afterglow discovery and to the possibility given to the optical and radio astronomers  to discover the GRB optical counterparts, the long-time mystery about the origin of these events has been solved. Now we know that GRBs are huge explosions, mainly ultra relativistic jets, in galaxies at cosmological distances. Starting from the first GRB detection with the Vela satellites, I will review the story of these discoveries, those obtained with \sax, the contribution to GRBs by other satellites and ground experiments, among them being {\venera, \cgro, \hete, \swift, \fermi, \agile, \magic, \hess,} which were, and some of them are still, very important for the study of GRB properties. Then, I will review the main results obtained thus far and the still open problems and prospects of GRB astronomy.}

\begin{document}


\section{Introduction}
\label{intro}
{Gamma} 
 ray bursts (GRBs) are among the most intriguing phenomena in the Universe. They are sudden bright flashes 
of celestial gamma-ray radiation, with~variable
duration from milliseconds to several hundreds of seconds. Sometimes, their duration achieves 
thousands of seconds. Most of their emission extends from a few keV up to tens of MeV, but~also GeV emission has been detected in many of them, and, in~some cases, also TeV emission.
With the current X-ray/gamma-ray detectors, their occurrence rate is 2--3 per day over the entire sky.
Their arrival times are unpredictable {as is their arrival direction.} When they occur, their brightness outshines any other celestial X-ray/gamma-ray~source. 

In this paper I will review the main steps of the research studies on GRBs. In~particular, in~Section~\ref{discovery}{} 
 I will start with the GRB discovery with the \vela\ satellites. In~\mbox{Sections~\ref{main efforts} and~\ref{Batse era}} I will discuss the main efforts performed soon after the GRB discovery, mainly with \venera\ satellites, the~BATSE experiment and its main results. In~Section~\ref{afterglow discovery} I will discuss the peculiar story of the \sax\ discovery of the GRB afterglow and the determination of their extragalactic origin. In~Section~\ref{immediate consequences} I will review the immediate consequences of the \sax\ discovery on the scientific community and on GRB theoretical models, while in~Section~\ref{other sax discoveries} I will summarize the main other results obtained with \sax\ on GRBs.~Section~\ref{post-sax era} is devoted to discussing the main results obtained in the post-\sax\ era.  Theoretical aspects, like GRB progenitors and the inner engine, are discussed in Sections~\ref{s:progenitors} and \ref{engine-physics}. The~key role played by GRBs for the birth of the multi-messenger era is reviewed in Section~\ref{multimessenger}, while, in~Section~\ref{future}, the~still open issues about GRBs and the importance of GRBs for cosmology and fundamental physics are summarized. In~the last two \mbox{Sections~\ref{s:future missions} and \ref{s:astena},} I will discuss future prospects for GRB astronomy, particularly the space and ground facilities that are expected to solve many of the still open issues of this new and very intriguing research~field.

\section{GRB~Discovery}
\label{discovery}

The earliest GRBs were discovered by chance in 1967 with the American {\em Vela} satellites. These satellites were a series of 12 military spacecraft with a life time of the order of \mbox{1 year,} launched between 17 October 1963 ({\em Vela 1A} and {\em 1B}) and 8 April 1970 ({\em Vela 6A} 
 and {\em 6B}). The~goal of these satellites was to monitor the explosion of nuclear bombs in the terrestrial atmosphere, which were vetoed by the ``Partial Test Ban Treaty'' issued on 10 October 1963. In~particular, with~these satellites, the territory of the Soviet Union was monitored, which was the main nuclear-capable state that could  perform these tests in the atmosphere.  The~satellites were spinning and had on-board gamma-ray detectors. The~most sophisticated instrumentation was aboard {\em Vela 5} and {\em Vela 6}. \textls[-25]{It included a gamma-ray experiment which consisted of six CsI scintillation crystals, with~a total volume of 60 cm$^3$,  distributed so to achieve nearly isotropic sensitivity. The~energy  passband was~0.2--1~MeV in the case of {\em Vela 5} and 0.3--1.5 MeV in the case of {\em Vela 6} \citep{Klebesadel73}. The~scintillators were shielded for charged particles, thanks to a shield made of high atomic number materials. The~time resolution of the data collected from these satellites was continuously improving, from~32~s integration time for {\em Vela 1} to 64~ms for {\em Vela 5} and {\em Vela 6}.}

\begin{figure}[H]
\includegraphics [width=0.45\textwidth]{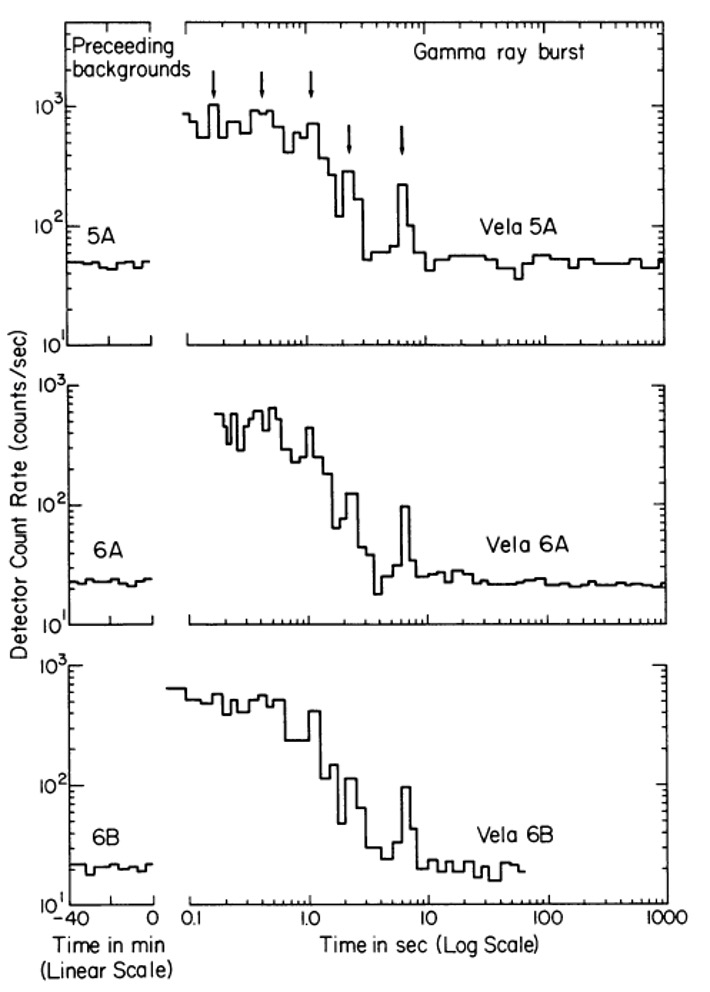}
\caption{One of the 73 gamma-ray bursts observed with the {\em Vela} satellites. It shows the count rate as a function of time for the gamma-ray burst of {22 August 1970}.
~Arrows indicate some of the common structures. Background count rates immediately preceding the burst are also shown.  Reprinted from~\cite{Klebesadel73} © AAS. Reproduced with permission.}
\label{f:Klebesadel1973}  
\end{figure}

On 2 July 1967, the~{\em Vela 3} and {\em Vela 4} satellites detected a flash of gamma-ray emission unlike any known nuclear weapon signature. Other similar events were observed with other {\em Vela} satellites launched later and with better instruments. By~analyzing the different arrival times of the bursts as detected by different satellites, the~Los Alamos team led by Ray Klebesadel was able to determine rough estimates for the sky positions of sixteen bursts and to definitively rule out a terrestrial or solar origin. The~discovery was eventually published in 1973 \citep{Klebesadel73} with the title ``Observations of Gamma—Ray Bursts of Cosmic Origin''. In~Figure~\ref{f:Klebesadel1973}, one of the GRBs observed with the \vela\ satellites is shown. In~the time period of 1969 to 1979, the \vela\ spacecraft (5 and 6) recorded 73 gamma-ray bursts. A~preliminary catalog of  events was reported by~\cite{Strong74}.

%
%
\vspace{-6pt}

The first questions were as follows: Which are the sites of these events? Which are their progenitors? Which is the power released in these events? The solution to these issues implied the solution of complex observational problems, like an accurate localization of the events in order to associate them to a possible known source. This was  a tough task in the gamma-ray energy band, where the source position could not be accurately determined. A~possible solution was the search for GRB counterparts at longer~wavelengths.

\section{Main Efforts Soon after the GRB~Discovery} 
\label{main efforts}

Many satellite missions in the 1970s and 1980s (e.g., the~Russian satellites \emph{Venera} 11, 12, 13, 14, Prognoz 6, Prognoz 9, Konus, Granat, and the~American Pioneer--Venus Orbiter and Solar Maximum Mission) were launched with instruments  also devoted to detect GRBs (see~\cite{Niel76,Barat81}). Relevant results about their origin were obtained with the Venera satellites. Aboard \venera\ 11 and \venera\ 12, there was a hard X-ray/soft gamma-ray GRB experiment, \konus, developed by the Ioffe Physico--Technical Institute in St. Petersburg, which consisted of six NaI(Tl) scintillator detectors, which were completely open, apart from a shield on the sides and bottom. The~detectors were oriented along six different directions and covered all the sky. The~different orientation of the detector axes allowed to obtain a localization of the GRB sources with an accuracy $\ge 4$ deg, while, when the mutual distance of the \mbox{two satellites} was also taken into account, a~localization of the GRB direction in the arcminute range was even possible through triangulation. In~addition to the localization, it was also possible to obtain the temporal structure and photon spectrum of the events \citep{Mazets79,Mazets81b}.   A~modified version of the \konus\ experiment was also flown aboard \venera\ 13 and \venera\ 14 launched in 1981 October 30 and November 4, respectively. The~main differences concerned the number of energy channels and a better time resolution. For~example, the~temporal accumulation of the photon spectra was 0.5 s instead of 4~s.

\begin{figure}[H]
\includegraphics [width=0.80\textwidth]{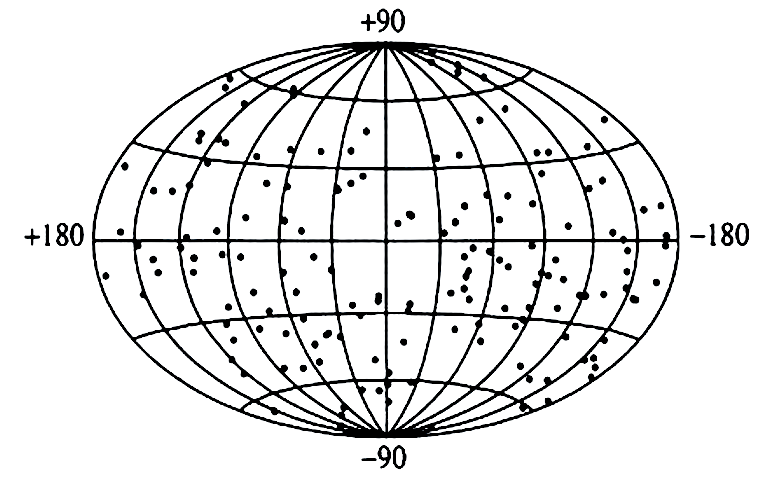}
\caption{Distribution of GRBs in the sky in Galactic coordinates as obtained with the Konus experiment  aboard the \emph{Venera} 11--14 missions. Reprinted from~\cite{Mazets88a}.}
\label{f:skydistr-Mazets88}  
\end{figure}

The \konus\ results on GRB were very interesting. A time-resolved correlation was found between the source intensity and peak energy (interpreted as bremsstrahlung temperature) of the $E F(E)$ spectrum, where $E$ is the photon energy and $F(E)$ is the GRB energy spectrum. In~addition, the earliest evidence was found of an isotropic distribution of the GRB positions in the sky~\cite{Mazets81b,Mazets88a} as shown in Figure~\ref{f:skydistr-Mazets88}. This distribution could mean that the GRBs are either very close to the Earth or very~far. 

%
%
\unskip

\section{The BATSE~Era}
\label{Batse era}

\begin{figure}[H]
\includegraphics [width=0.55\textwidth, angle=90]{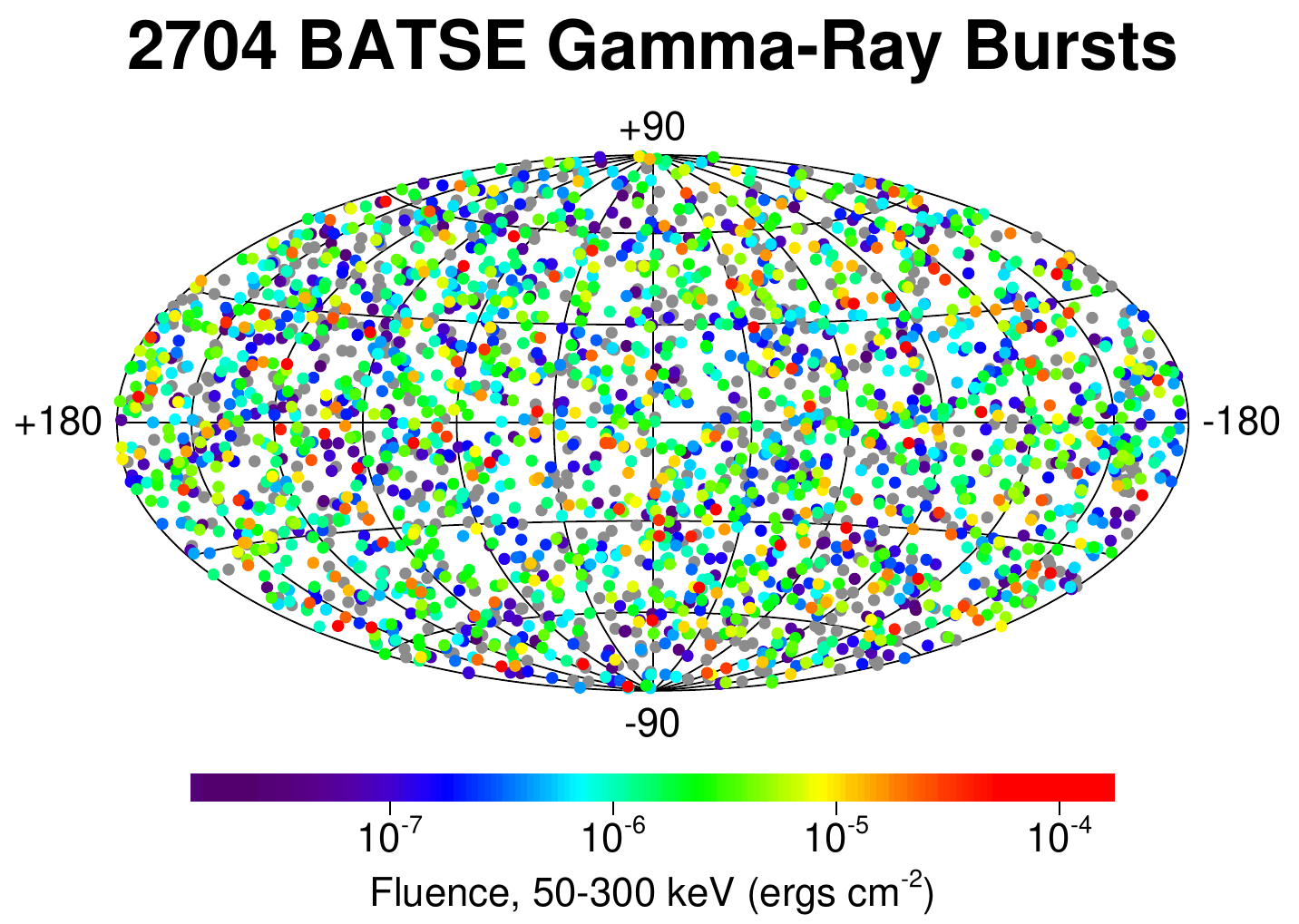} 
\caption{{Sky} 
 distribution in galactic coordinates of 2704 GRBs detected with the BATSE experiment aboard {the} 
 \cgro\ satellite. Adapted from~\cite{Paciesas99} © AAS. Reproduced with permission. ({Credits:} 
 \url{https://gammaray.nsstc.nasa.gov/batse/grb/skymap/} 3 May 2024).}
\label{f:skydistr-Mazets88-Paciesas99}  
\end{figure}

\begin{figure}[H]
%
\includegraphics [width=0.55\textwidth]{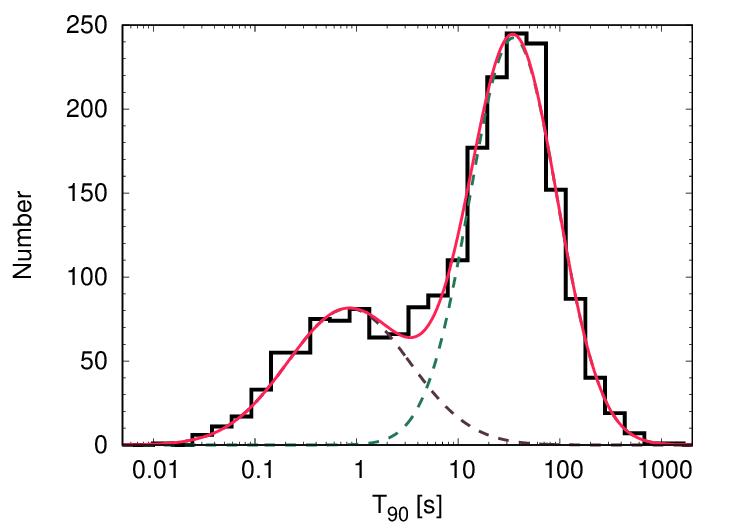} 
\vspace{-3pt}
\caption{{Distribution} 
 of $T_{90}$ for 2006 GRBs detected with BATSE with significant $T_{90}$ estimates. $T_{90}$ is defined as the time during which the GRB cumulative counts increase from 5\% to 95\% of the total detected~counts. Dashed lines give the best fit to the $T_{90}$ distribution.}
\label{f:bimodality}  
\end{figure}

\begin{figure}[H]
\includegraphics[width=0.8\textwidth]{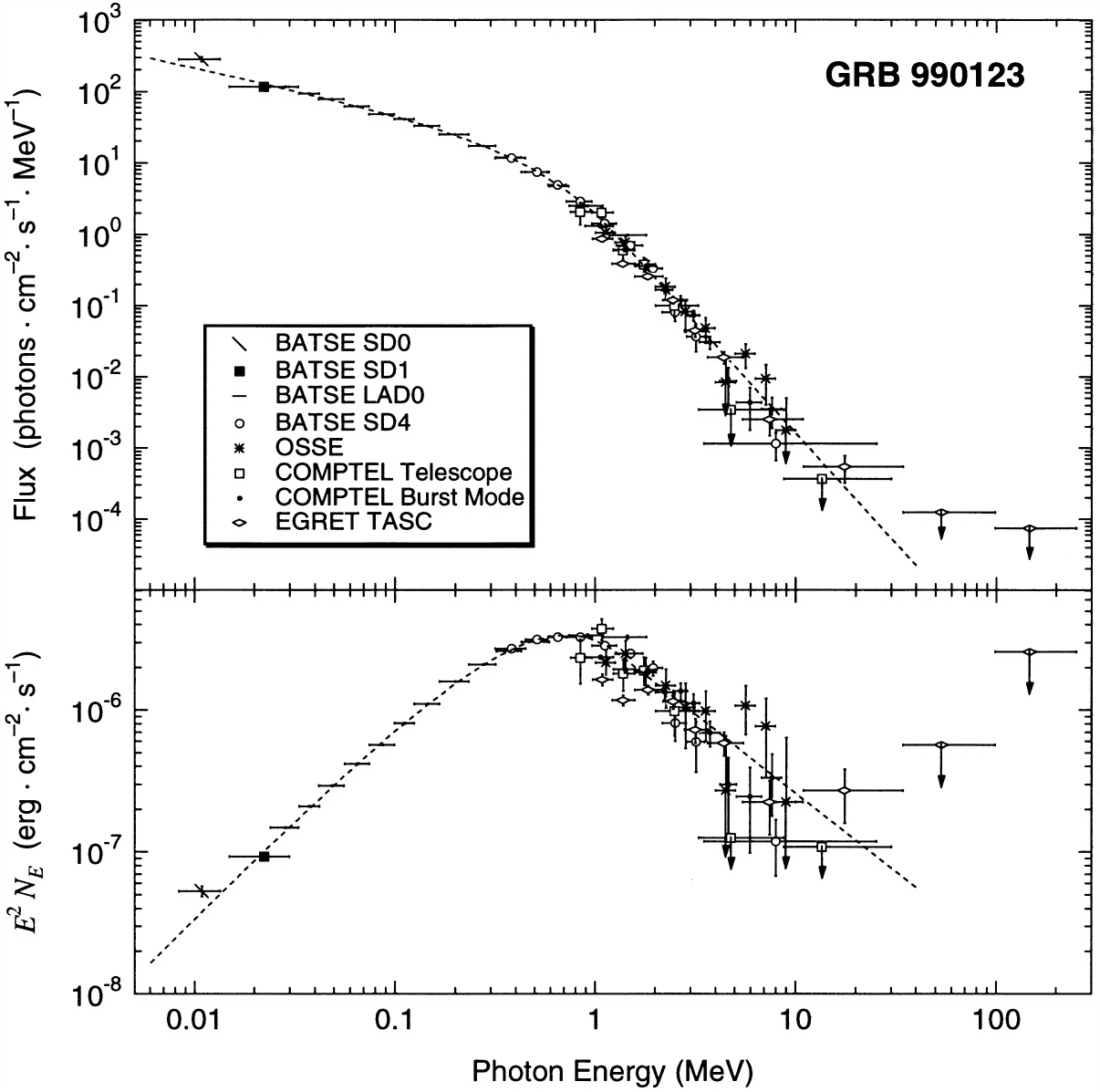}
\caption{{An example} 
{of GRB} 
 \textls[-15]{photon spectrum and the corresponding $E^2 N(E)$ function.  It shows the \batse\ spectrum of the GRB occurred on 1999 January 23 (GRB\,990123), that was  also detected and promptly accurately localized with \sax. Reprinted from~\cite{Briggs99} © AAS. Reproduced with~permission}.}
\label{f:GRBspectrum}
\end{figure}

\begin{figure}[H]
\includegraphics[width=0.6\textwidth]{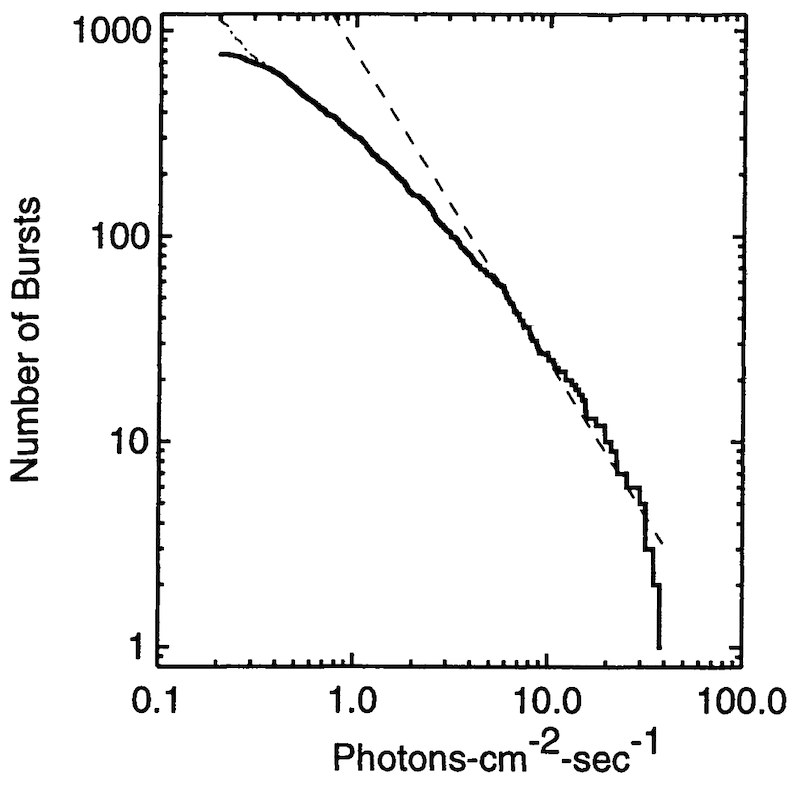}
\caption{{Integral} 
 $\log N$--$\log P$ distribution of 772 GRB detected {with} 
 \batse\ in the range \mbox{50--300 keV} band.
The peak flux $P$ is integrated over 1.024 s. The~dot-dashed line gives the correction for instrumental trigger efficiency,
while the dashed line gives the power-law slope ($-3/2$) expected in the case of an homogeneous distribution of GRBs in space.
Reprinted from~\cite{Meegan96}.
}
	\label{f:intensity distribution}
\end{figure}

The definitive answer to the GRB distribution in the sky was given by the {{BATSE} 
} (Burst And Transient Source Experiment) experiment aboard the American satellite \cgro. Their isotropy in the sky was definitely confirmed (Figure~\ref{f:skydistr-Mazets88-Paciesas99}).
%
%
The \cgro\ mission (e.g., \cite{Gehrels93}), designed in the 1980s, was launched on 5 April 1991 by the Space Shuttle Atlantis and reentered the Earth atmosphere on 4 June 2000.
Three hard X-ray/soft gamma-ray experiments were on board: an Oriented Scintillation Spectrometer Experiment ({OSSE}), a~COMPton TELescope ({COMPTEL}), and~\batse. This experiment, whose PI was Gerald J. Fishman, a scientist of the NASA Marshall Space Center in Huntsville (see~\cite{Fishman92}), was developed mainly to detect and locate GRBs along with the study of their temporal and spectral properties. It consisted of eight  completely open NaI(Tl) Large Area Detectors (LADs) at the
corners of the spacecraft, each sensitive in the 30 keV--2 MeV energy range and with an exposed area of 2025 cm$^2$. 
For each LAD, there was a smaller spectroscopy detector (SD) with a detection area of about 600~cm$^2$ \citep{McNamara93}, optimized for energy resolution and broad energy coverage (10~keV--11~MeV).
The GRB shape could be transmitted with different time resolutions, down to $\upmu$s time scales.
\batse\ was capable of establishing not only the isotropic distribution of GRBs in the sky, but~also many other properties of the GRB prompt emission, like the bimodality of the GRB prompt emission duration $T_{90}$ (see Figure~\ref{f:bimodality}), the~spectral distribution of the prompt emission (see Figure~\ref{f:GRBspectrum}) described by a smoothed broken power law (also called the Band function from the paper by Band~et~al.~\citep{Band93}, who proposed it), and the~GRB intensity distribution (see Figure~\ref{f:intensity distribution}).
%
%

\vspace{-6pt}
\unskip

%
\unskip

%
%

In spite of these and many other results, due to the uncertainty in the \batse\ GRB localization capability (several degrees), all the numerous efforts to discover  X-ray or optical or radio counterparts of the \batse\ GRBs were unsuccessful, and~the sites of the GRBs were not~discovered.

The uncertainty in the GRB sites, and thus in their distance scale, was also a matter of debate as~demonstrated by the meeting that took place at the Baird Auditorium of the Smithsonian museum of Natural History in Washington DC (USA) on 22 April 1995, where there were two main points of view about the GRB sites and progenitors, one led by Bodhan Paczynski and the other led by Don Lamb. The~viewpoint of Paczynski was the following~\cite{Paczynski95}: ``At this time, the~cosmological distance 
scale is strongly favored over the Galactic one, but~is not proven. A~definite proof (or dis-proof) could 
be provided with the results of a search for very weak bursts in the Andromeda galaxy (M31) with an 
instrument ~10 times more sensitive than \batse. If~the bursters are indeed at cosmological distances then 
they are the most luminous sources of electromagnetic radiation known in the Universe. At~this time we have no clue as 
to their nature, even though well over a hundred suggestions were published in the scientific~journals''.
 
Instead, Don Lamb had a completely different point of view~\cite{Lamb95}: ``We do not know the distance scale to gamma-ray bursts. Here I discuss several observational results and theoretical calculations which provide evidence about the distance scale. First, I describe the recent discovery that many neutron stars have high enough velocities to escape from the Milky Way. These high velocity neutron stars form a distant, previously unknown Galactic corona. This distant corona is isotropic when viewed from Earth, and~consequently, the~population of neutron stars in it can easily explain the angular and brightness distributions of the BATSE~bursts''.
  
Thus, in~1995, in~spite of some convictions, the~distance scale of the GRB sites was still an open issue, with~the following premonitory conclusion of the debate by Martin Rees~\cite{Rees95}: ``I'm enough an optimist to believe that it will only be a few years before we know where (and perhaps even what) the gamma-ray bursts~are''.


\section{The \sax\ Afterglow~Discovery}
\label{afterglow discovery}

The solution of the mystery about the GRB sites was obtained with the Italian \sax\ satellite with Dutch participation~\cite{Boella97b}: in seven months from the beginning of its operational phase (October 1996), thanks to \sax, it was possible to establish that GRBs are huge explosive events in galaxies at cosmological distances.  Why did \sax\ solve the mystery of the GRB distance? This story merits description, recalling the evolution of the \sax\ science goals and its~design.

\subsection{{\em~SAX} Initial Goals and Evolution in Its Configuration}

 Actually, the initial main goals of this satellite, initially named SAX (Satellite Astronomia X, in~Italian), had as its initial main goals the following: (1) the study  of celestial X-ray sources in a broad energy band (0.1--300 keV) with narrow field instruments; (2)~the X-ray (2--30 keV) monitoring of the sky, in~particular, of the galactic plane, with~wide field~cameras.

The reason for such science goals for SAX was that, when SAX was proposed and approved in 1981 by the National Space Plan (later, in~1988, Italian Space Agency), X-ray astronomy was 20-year aged, with~the best knowledge of the X-ray sky mainly obtained with satellite missions in the low energy band ($<$10~keV). The~maximum sensitivity and localization were achieved in the softer X-ray band ($<$3~keV) with the introduction of grazing incidence optics in satellite missions, like {\em Einstein} \citep{Giacconi79}. The~celestial hard X-ray band ($>$20~keV) was explored mainly with balloon-borne experiments, and~source spectra were well measured only for the strongest sources. With~the growing in the sensitivity, the~source spectra appeared more complex, and a source variability was detected at all time scales and at all energies. But~a systematic study of source time variability, along with the covering of a broad energy band  with high sensitivity, was still not performed. Thus, the main target of SAX was that of covering, with~narrow field instruments, an~unprecedented energy range with a sensitivity as balanced as possible, and, at~the same time,~monitoring wide portions of the sky for variability studies. 
GRBs were only mentioned as possible X-ray targets  for wide field cameras (see below) for the possible affinity of GRBs with emission from neutron star binaries, on~the basis of the discovery, with~the \emph{Venera} 11 satellite, of~the famous flaring source in Dorado firstly observed on 5 March 1979 \citep{Mazets79b} but~later recognized as a different class of sources (Soft Gamma-Ray Repeaters or magnetars for their very strong magnetic field).  

To achieve the mentioned goals, the~initial SAX payload proposed included three instruments, two with a narrow field of view (FOV) and one with a wide FOV.
The narrow field instruments (NFIs), oriented in the same direction, were as follows:
\begin{itemize} 
\item
A Gas Scintillator Proportional Counter (GSPC) with 2--35 keV energy passband, surmounted by a coded mask (3 deg FOV) with arcminute imaging capability. The~Principal Investigator (PI) was Giuseppe Manzo from the CNR (Italian National Research Council) Institute of Cosmic Physics and Informatics.
\item
A Phoswich Detection System (PDS) at higher energies (15--300 keV), consisting of four independent detection units, each one made of a sandwich of NaI(Tl) plus CsI(Na) scintillator crystals. The~NaI(Tl) was used as the main detector, with the CsI(Na) as an active shield from the bottom. This technique, called phoswich ($=$PHOSphor sandWICH), had been demonstrated to provide a very low instrument background (BKG). To~further reduce the BKG, four slabs of CsI(Na) detectors, in~anti-coincidence with the four phoswich units, laterally covered the instrument. The~PDS FOV, of~1.5 deg (Full Width at Half Maximum, FWHM), was obtained by means of honeycomb collimators. The~PI was myself, at~that time a scientist of the CNR Institute of Technology and Study of the Extraterrestrial Radiations. See Frontera~et~al. \citep{Frontera97}.
\end{itemize}

The wide field instrument consisted of two wide field cameras (WFCs), with~axes perpendicular to the NFI and in opposite directions to each other. Each camera was made of a proportional counter surmounted by a coded mask. The~energy band of each WFC was 2--28 keV, and its FOV was $20^\circ \times 20^\circ$ (FWHM) with imaging capability with an angular resolution of 3--4 arcmin. The~PI was Rick Jager from the Institute of Space Resarch (SRON) in Utrecht (Holland). See Jager~et~al. \citep{Jager97}.

The rationale to have on board the same mission two WFCs, oriented along
two opposite directions perpendicular to NFIs, was to have the widest FOV for~the monitoring of a large number of variable sources, in~particular, new transients,  and~to have the possibility to perform shorter pointed observations at particular states of~them.

During the {\em industrial phase A} study performed by the AERITALIA (later, Alenia Spazio), the~GSPC was replaced by a set of four focusing telescopes having the same on-axis direction, with~one of them (Low Energy Concentrator Spectrometer, LECS) with a passband from 0.1 to 10 keV, and~the other three (Medium Energy Concentrators Spectrometers, MECS) with a passband from 1.3 to 10 keV. All of them had a FOV of 0.5 deg and an angular resolution of the order of 1 arcmin. 
The PI of LECS and MECS was Giuliano Boella from the CNR Institute of Cosmic Physics, with~responsibility of the LECS focal plane detector by Arvind Parmar from ESA-ESTEC, Noordwijk (The Netherlands). See Boella~et~al. \citep{Boella97a}, and Parmar~et~al.~\citep{Parmar97}. 

Given the high scientific interest at that time for the cyclotron line features in the spectra of X-ray pulsars, the~GSPC instrument, for~its very good energy resolution, was still included in the payload with no coded mask and a higher (5 atm) gas pressure (HPGSPC). Its final FOV was 1.1° FWHM, and its energy passband from 4 to 60 keV. See Manzo~et~al.~\citep{Manzo97}. 

The original proposal did not include GRBs as a main science goal. This was due to the fact that in the early 1980s, our knowledge of GRBs was strictly confined to gamma-rays or very hard X-rays. It was not clear whether an X-ray phenomenology could be expected. For~sure, until~that time, no transient X-ray detected phenomenon had been associated with a GRB. Moreover, due to the wide spread of interpreting models, it was not clear whether any X-ray delayed emission could be~detected. 

\subsection{Addition of a Gamma-Ray Burst Monitor {{GRBM}} and Establishment of a Team for GRBs Identification and Localization}
In 1984, during~the phase A study mentioned above, I proposed to use the four slabs of CsI(Na) anti-coincidence detectors as~a Gamma-Ray Burst Monitor (GRBM) with a passband from 40 to 700 keV (see Internal Report of CNR-TESRE, No. 99, 1984).
The motivation for this implementation was that the axis of two of these shields was parallel to that of the two WFCs. This feature was very suggestive. Indeed, we expected that about three GRBs per year (see SAX Observers' Handbook, Issue 1.0, 1995) could enter into the common field of view of WFCs and GRBM, and thus they could be identified as true GRBs by GRBM and localized by WFCs within 3--5 arcmin, an~accuracy never achieved at that time by the GRB monitors already launched or only designed. The~GRB identification with GRBM was crucial because~transient events detected with only WFCs could be originated by other phenomena (e.g., star flaring and X-ray bursts), while their contemporary detection in the GRBM energy band ($>$40 keV) was an almost certain signature of being true GRBs. Obviously, it was required  to develop a proper electronic chain, an~in-flight trigger system, and~a prompt download of the data. The GRBM proposal had also an international resonance \citep{Hurley86}.

\begin{figure}[H]
\includegraphics[width=0.5\textwidth, angle=0]{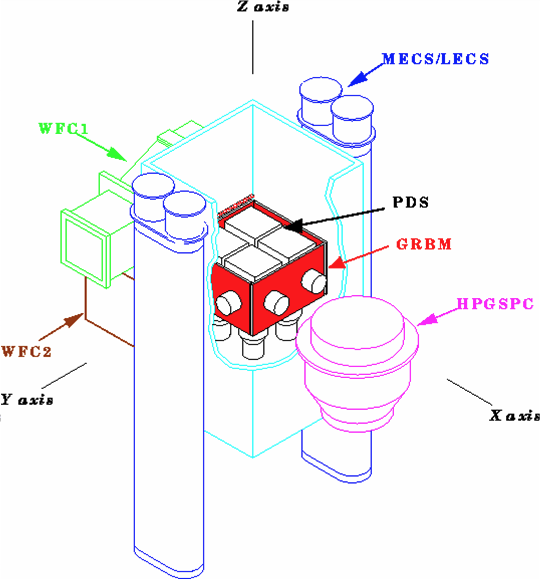}
\caption{{The} 
 final \sax\ payload. \grbm\ is shown in red as part of the PDS instrument. It is made of the four independent slabs of CsI(Na) scintillators, 1 cm thick. They were initially foreseen as active anticoincidence shields of the 4 phoswich units. Notice that two opposite GRBM units were oriented as the two \wfcs.}
\label{f:saxpayload}
\end{figure}

In 1990, a~cheaper version of the GRBM proposal was approved by the Italian Space Agency (ASI) as a further instrument of SAX. With~this addition, the~final SAX payload  became  that shown in Figure~\ref{f:saxpayload}.
Once the GRBM proposal was approved, its design was better defined and improved, in~the context of limited resources, and~without interfering with the anti-coincidence function of the CsI(Na) slabs, which had been the primary driver of the PDS detector design. For~a full  GRBM description, see Ref.~\cite{Costa98}.

%
%
By means of a Monte Carlo code, we found that the different orientations of the GRBM units could be exploited for obtaining a crude GRB localization that was sufficient for deconvolving the GRB count spectra \citep{Pamini90}. The~implementation of an optimized  response function of GRBM  required a very detailed description of the entire SAX satellite, which was obtained with both simulations and ground calibrations. The~simulations were performed thanks to the development of a very detailed Monte Carlo code \citep{Rapisarda97}, while the ground calibrations were performed at different steps, with~the last one having a calibration campaign performed at ESTEC (Noordwijk, The Netherlands) after the integration of the instrument in the satellite~\cite{Amati98}.

Before the SAX launch, in~response to an international call by the SAX collaboration, the~established PDS/GRBM team submitted a proposal to obtain WFCs data in the case of GRBs identified with GRBM. The~goal was to promptly and accurately localize GRBs and perform a follow-up of the localized event with SAX Narrow Field Instruments (NFIs). Also, the BATSE team submitted a similar proposal for GRBs identified with the BATSE experiment. Both proposals were approved by the SAX Time Allocation~Committee.

\subsection{The SAX Launch and the First Detected~GRBs}

\begin{figure}[H]
\includegraphics [width=0.45\textwidth, angle = 0]{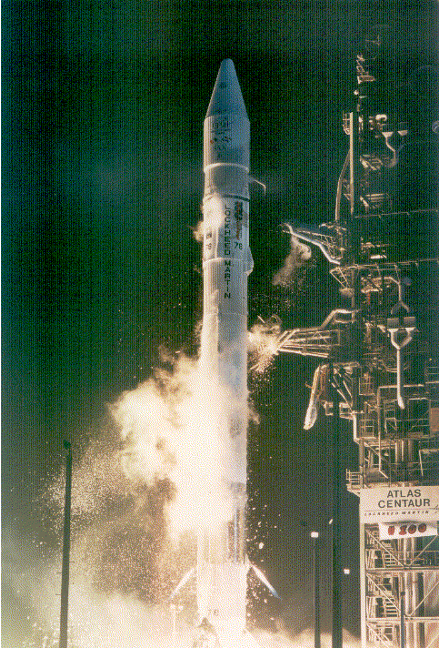}
\includegraphics [width=0.45\textwidth, angle = 0]{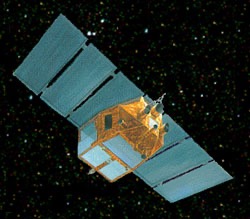}

  \caption{(\textbf{{Left}
}): SAX in the Cape Canaveral launching pad before~its launch with an Atlas-Centaure double rocket. (\textbf{{Right}}): an artistic view of the satellite in~flight.}
    \label{f:sax-launch}
\end{figure}

\textls[-15]{SAX was launched on 30 April 1996 from Cape Canaveral with an
Atlas-Centaure double rocket, reaching an altitude of 600 km with a final orbit inclination of 3.8 deg (see Figure~\ref{f:sax-launch}). Soon after the successful launch, the~satellite name was modified, becoming \sax\, in~honor of the famous Italian physicist Giuseppe (called Beppo by his friends) Occhialini (see \hl{} 
~\url{https://en.wikipedia.org/wiki/Giuseppe_Occhialini} (accessed on 2 May 2024)).}
%
%

The commissioning phase had a two month duration up to the end of June 1996, while the Science Verification Phase (SVP) had a
3-month duration  
(July--September 1996). The~satellite started its operational phase in October 1996. The~telemetry link (10 min per orbit) with the satellite was obtained by means of the ASI ground station in Malindi, which allowed to download the mass memory with the satellite data  and to up-link~telecommands. 

The first GRB that we identified with GRBM and found in the field of view of one of two WFCs, occurred  during the SVP phase, on~20 July 1996. However, it could be accurately localized only 20 days after the event time, and, thus, after~about one month, the~\sax\ NFIs were pointed to the GRB direction, with~no detection of an X-ray counterpart~\cite{Piro98a}.
 
From this result, it was clear that a possible residual X-ray radiation, if~any, could have been found only in the case that it was possible to point, as~soon as possible, the~NFIs along the direction of well-localized events. To~this end, we optimized all the  procedures to minimize the time needed to establish whether a GRBM event was a true GRB, whether a contemporary source image was detected by WFCs, and, in~the positive, to submit the request of a prompt \sax\ Target of Opportunity (TOO) to re-point the NFIs toward the localized~event. 

The first GRB detected and localized with this procedure occurred on 11 January 1997. The~light curve of this event (GRB~970111) is shown in the top panel of Figure~\ref{f:970111-errorbox}. Its earliest position was derived with a 10 arcmin error radius (see bottom panel of Figure~\ref{f:970111-errorbox}). On~the basis of this result, an~X-ray follow-up with NFIs was started 16 hrs later. The~same derived position was given to our collaborators for the search of an optical/radio counterpart. In~the WFC error box, Dale Frail and his colleagues, with~the VLA (Very Large Array) radio telescope in Socorro (NM, USA), observed an unusual radio source (VLA1528.7 + 1945), variable on time scales of days.  A~similar variability was never found before in a radio source. Thus, the source was assumed to be the likely radio-counterpart of the GRB event, and~a paper was promptly submitted to the {\em {Nature} 
} journal. However, after~ about 20 days, with~a better deconvolution function, we obtained a more accurate WFC position and error box. The~error radius became only 3 arcmin, and the centroid position was about 4 arcmin far from the previous one (see bottom panel of Figure~\ref{f:970111-errorbox}). The~radio source was no more consistent with the new WFC position. Similarly, the~two bright X-ray sources found with the \sax\ NFIs \citep{Feroci98}, with~one of them coincident with the radio source position, were not compatible with the new WFC position. The~ paper submitted to \emph{Nature} was withdrawn and later published in \emph{ApJ}~\cite{Frail97a}.

 In the current time of multi-messenger astromony (see later), the~experience acquired with GRB~970111 is an important teaching. A~very accurate positional coincidence is crucial for associating a GRB event to an optical/IR counterpart or to a gravitational wave (GW) signal or to a neutrino event, and~viceversa. A~mission concept like \theseus~(see later), thanks to its accurate capability to localize a GRB  in X-rays and to perform a prompt follow-up with the infrared telescope onboard, appears to be the best solution for the prompt association of a GRB event to an optical counterpart. Also, the association of a gravitational wave (GW) signal or a neutrino event to a GRB should follow similar~strategies. 
 
 \vspace{-6pt}

%
%
\begin{figure}[H]
\includegraphics [width=0.75\textwidth, angle = 0]{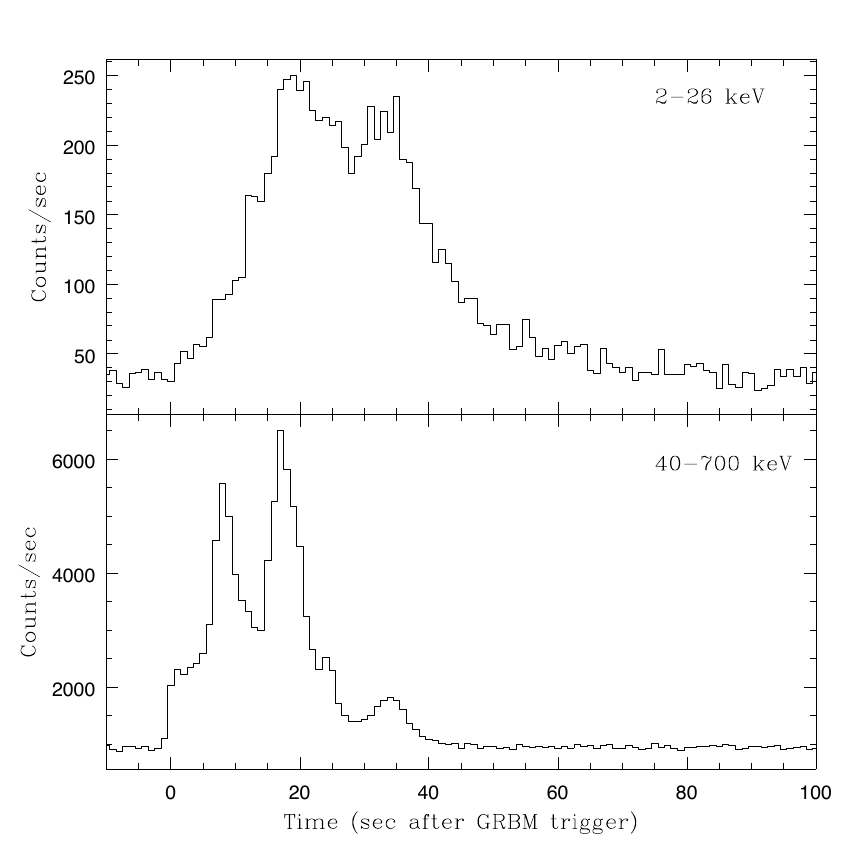}
  \caption{\emph{Cont}.}
    \label{f:970111-errorbox}
\end{figure}

\begin{figure}[H]\ContinuedFloat
\includegraphics [width=0.75\textwidth, angle = 0]{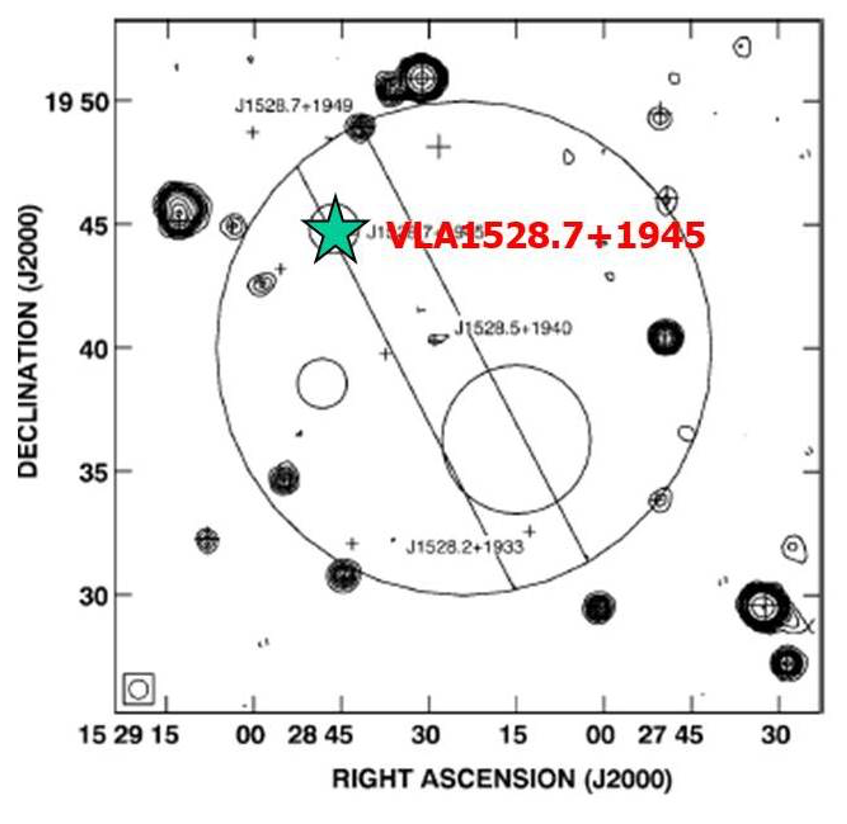}
  \caption{(\textbf{Top panel}): {Light} 
 curve of GRB~970111. Reprinted from~\cite{Feroci98}. (\textbf{Bottom panel}): The largest circle gives the earliest \wfc\ error box, while the smaller circle gives the latest \wfc\ error box. The~strip crossing the largest circle is the error box derived with the Interplanetary Network (IPN). As~can be seen, in~the intersection of the refined \wfc\ error box with the IPN annulus, no source was observed. Adapted from the figure of~\cite{Frail97a}.}
    \label{f:970111-errorbox}
\end{figure}

\subsection{The First GRB Afterglow~Discovery}

\begin{figure}[H]
\includegraphics[width=0.68\textwidth, angle=0]{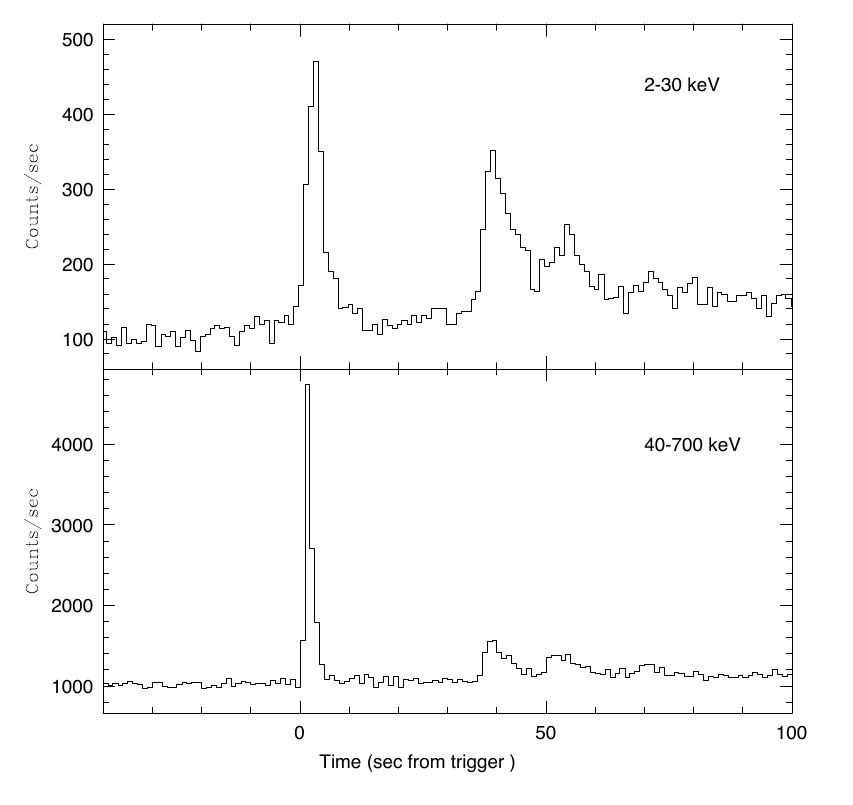}\\
\includegraphics[width=0.78\textwidth, angle=0]{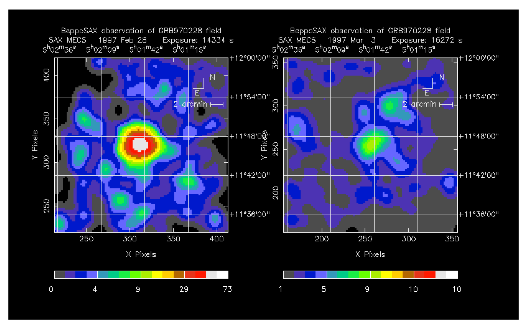}
\caption{(\textbf{Top panel}): {Light} 
 {curve} 
 {of} 
 GRB~970228 as detected with \wfc\ and \grbm. (\textbf{Bottom panel}): X-ray images of GRB~970228 as detected with MECS telescopes at two different times. The~image on the left is that obtained  8 to 16 h after the burst. The~image on the right is that obtained in the observation 3.5 days after
the burst. From~the first to the second X-ray observation, the~source had faded by a factor 20. Reprinted from~\cite{Costa97}.}
\label{f:GB0228}
\end{figure}

The mystery about the GRB sites was solved with the discovery of the X-ray and optical counterpart of GRB~970228. This GRB, identified with GRBM and localized with one of the two WFCs, showed a light curve with a bright peak, followed by a train of three more peaks of decreasing intensity (see top panel of Figure~\ref{f:GB0228}).

The NFI follow-up was performed 8 h after the time of the event and lasted about 9 h. In~the MECS FOV, a previously unknown  source (SAX~J0501.7 + 1146) was found with a flux of $(2.8\pm 0.4)\times 10^{-12}$~erg cm$^{-2}$s$^{-1}$ in the 2--10 keV band. The~source was pointed out again three days after, and it was found to have faded by a factor 20 (see bottom panel of Figure~\ref{f:GB0228}).

By subdividing the first MECS observation into three subsets, the~time behavior of the discovered source was derived. It resulted that the source was fading according to a power law $N(t)\propto t^{-1.33}$, where $t$ is the time since trigger.
Also, the flux detected by the \wfcs\ during the GRB tail of the light curve, in~the same energy band of MECS,  was found to be consistent with the fading law measured with NFIs. This fact is crucial to state that the X-ray source found was the delayed radiation (i.e., the~{\bf afterglow}) of the GRB event \citep{Costa97}.

In parallel, first the GRB coordinates obtained with WFC and later those obtained with NFIs, were given by the GRBM team 
to a few European Observatories working in service, and, in~parallel, distributed through IAU circulars (see~\cite{Costa97a}). Various observers performed
optical~observations. 

\begin{figure}[H]
  \includegraphics[width=0.8\textwidth, angle=0]{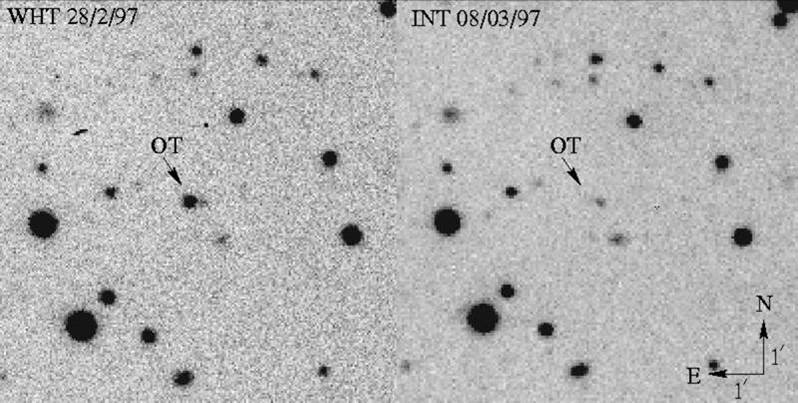}
\caption{Discovery of the optical counterpart of 
GRB~970228  at two different times (28 February 1997, and~8 March 1997). The~source is clearly fading. Reprinted from~\cite{Vanparadijs97}.}
 \label{f:fig970228-optical}
\end{figure}

Promptly, three optical observations of the distributed position were performed by the Dutch group led by Jan Van Paradijs. With~the same filter,  the~first observation was performed with the William Herschel Telescope on 28 February, and~the other two observations were performed on 8 March, one with the William Herschel  and the other with the Isaac Newton Telescope. All of these observations showed the presence of a previously unknown optical  source that
was fading (see Figure~\ref{f:fig970228-optical}) from $V = 21.3$ to $V>$ 23.6 \citep{Vanparadijs97}.

\vspace{3pt}

%
%

\vspace{-6pt}

\begin{figure}[H]
\includegraphics[width=1.0\textwidth, angle=0]{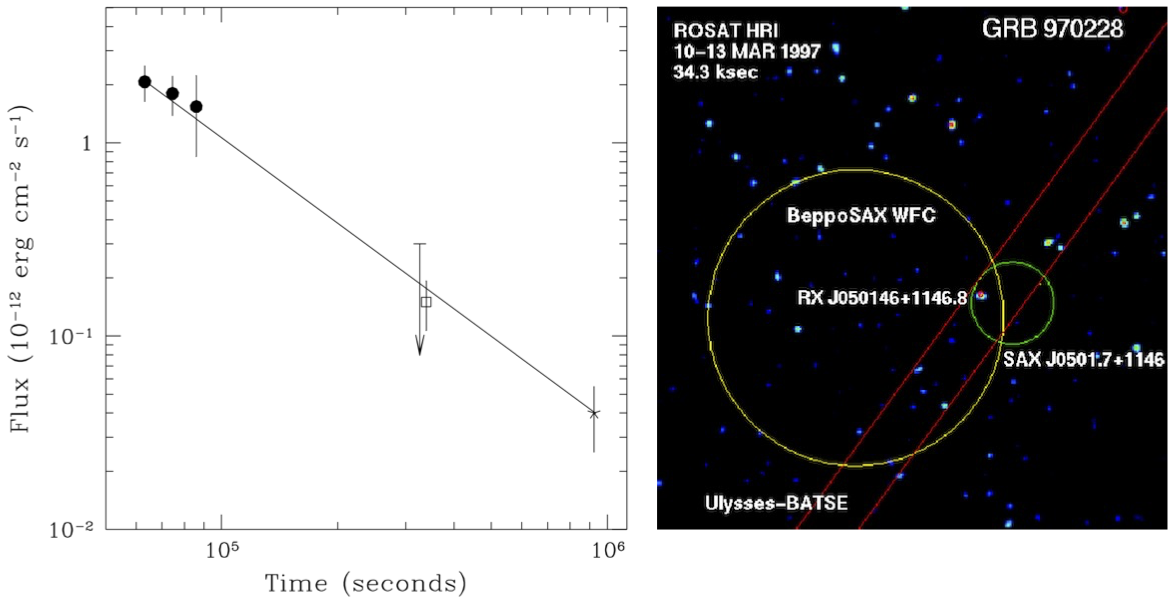}
\caption{(\textbf{Left}):{ Decline} 
 {in} 
 the GRB~970228 flux in 0.1--2.4~keV with time, starting from the follow-up time of GRB~970228 with \sax, uncorrected for galactic absorption. The~filled dots are the LECS data points, the~arrow is the LECS 3$\sigma$ upper limit,
the square gives the flux extrapolated from the MECS detection, and~the asterisk shows the \rosat\ HRI data point. The~best fit power-law decay (continuous line) is also shown. 
(\textbf{Right}): Image of the \rosat\ HRI FOV (8 arcmin wide) found during the GRB~970228 follow-up performed on 1997 March 10. 
Only one source (\mbox{RXJ050146 + 1146.8}) was found inside the error box (small circle with $\sim 50$ arcsec radius) of the fading source \mbox{SAX~J0501.7 + 1146.} In~addition, this \rosat\ source was coincident with the optical source (in red) within 2 arcsec. The~large circle shows the 3 arcmin error circle of GRB~970228 as determined with the \sax\ WFC, while the two straight lines give the uncertainty strip derived from the \sax\ GRBM and {\em Ulysses} timings of GRB~970228. Reprinted from Frontera~et~al. \citep{Frontera98b}.
}
 \label{f:rosat}
\end{figure}

Given the error box ($\sim$50 arcsec radius) associated with  SAX~J0501.7 + 1146, to~further confirm the association of this source with the GRB event, we proposed an observation with the High-Resolution Imager (HRI) aboard the X-ray satellite \rosat. This instrument, thanks to its sensitivity and high angular resolution (10 arcsec radius) in the energy band 0.1--2.4 keV, could provide a better position and a much lower error box of the source direction.
The observation was performed on March 10 and lasted three days. Eight sources were detected in the HRI FOV (20 arcmin) \citep{Frontera98b}, with~only one (RXJ050146 + 1146.9) in the error box of the X-ray source found with the \sax\ NFIs. The~source position (see right panel of Figure~\ref{f:rosat}) was coincident, within~2 arcsec,  with~the discovered optical transient~\cite{Vanparadijs97}. In~addition  its intensity was found to be fully consistent (see left panel of Figure~\ref{f:rosat}) with the extrapolation, at~the \rosat\ observing time, of~the power law decay estimated from the \sax\ LECS afterglow spectrum in the 0.1--2.4 keV energy band.   
With these results, the~association of the \rosat\ and \sax\ sources with the afterglow of GRB~970228 became~conclusive.

%
%
%

%
%

A detailed spectral analysis of GRB~970228 and its afterglow was also performed~\citep{Frontera98a}. While the spectrum of the prompt emission was consistent with the Band function and showed, within~each peak, a~hard-to-soft evolution, the~afterglow spectrum was a stable power law ($N(E) \propto E^{-2.04}$).  

Thus, as~a result of these X-ray observations, the temporal and spectral properties of the afterglow were in favor of a 
non-thermal process (e.g., \cite{Wijers97}) and became the basic building block for  GRB~theories.

\begin{figure}[H]
    \includegraphics[width=0.50\textwidth, angle=0]{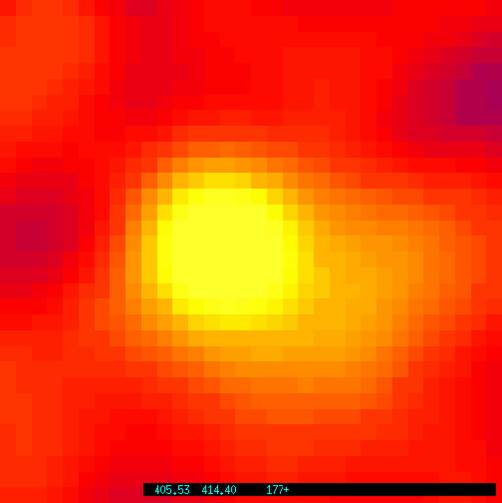}
\caption{Optical image of the GRB counterpart obtained with \hst, in~an observation performed 39 days after the burst. The~point-like source had further faded down to $V = 26.4$ and it seemed to be embedded in a faint nebular source with an extension of $\sim$1 arcsec  and a magnitude $V\approx 25$, likely, but~yet not yet necessarily, a~host galaxy. Figure provided by A. Fruchter, see~\cite{Fruchter97}.}
 \label{f:fig970228-hst}
\end{figure}

Thirty-nine days after the burst, a~further observation of the GRB~970228 optical counterpart was performed with the \hst. It was found that the point-like source had further faded down to $V =26.4$. In~addition, it appeared to be embedded in a faint nebular source (see Figure~\ref{f:fig970228-hst}) with $V\approx 25$ and an extension of $\sim$1 arcsec, likely, but yet not necessarily, a~host galaxy \citep{Sahu97}.
%
%
\unskip

\subsection{The First Measurement  of a GRB~Distance}

\begin{figure}[H]
    \includegraphics[width=0.98\textwidth]{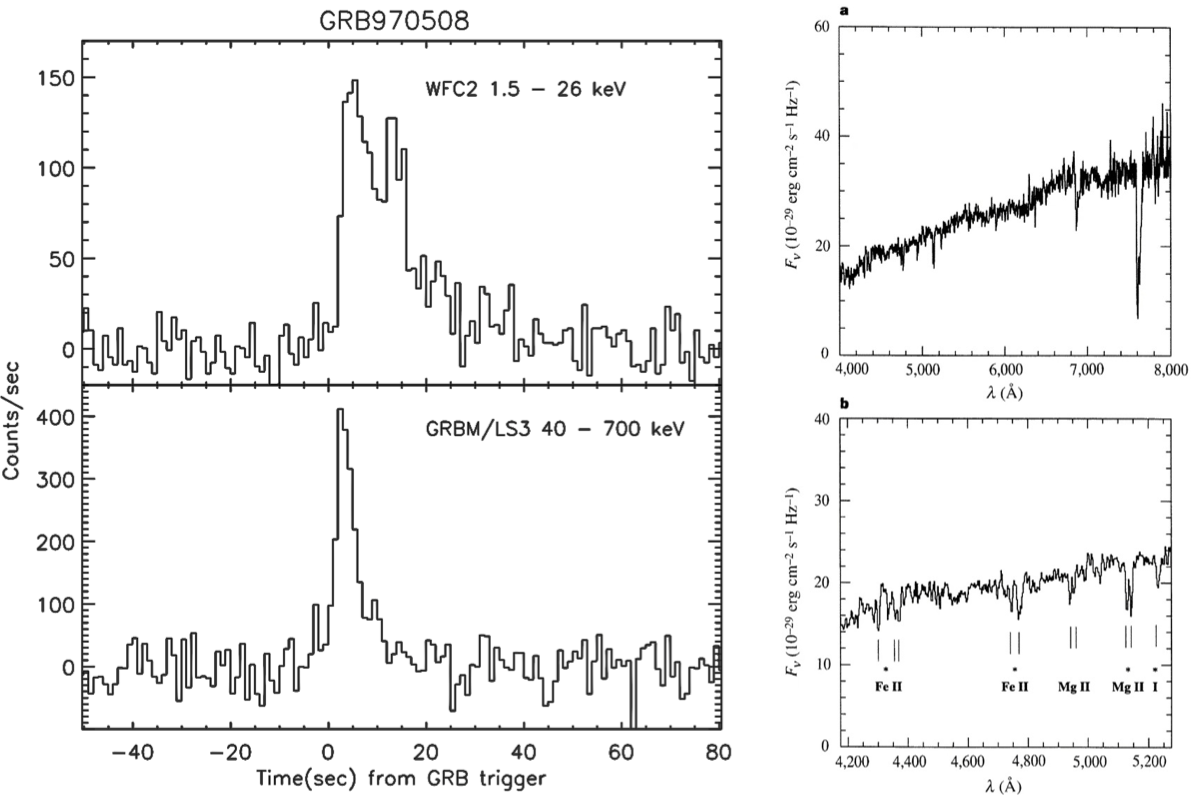}
\caption{(\textbf{Left top}): {Light} 
 {curve} 
 detected by WFCs (1.5--26 keV). (\textbf{Left bottom}): Light curve detected by GRBM (40--700 keV).  Reprinted from~\cite{Piro98b}. (\textbf{Right}): {Spectrum} 
 of the optical counterpart of GRB~970508 taken with the Keck Low
Resolution Imaging Spectrograph \citep{Metzger97}. ({\bf a}) Full spectrum; ({\bf b})~expansion of a limited region of the spectrum, with strong absorption lines. Various absorption lines were discovered: those marked with an asterisk correspond to a system at redshift $z= 0.835$, while the others at redshift of $z = 0.767$. The~largest $z$ values were found, weeks later, in~the emission lines from  the host galaxy associated with the fading object.} 
 \label{f:970508}
\end{figure}

The further turning point of the \sax\ discoveries occurred on {8 May 1997}, 
when GRB~970508 was identified with GRBM and localized with one of the two WFCs \citep{Costa97b,Heise97}. The~light curve of the event as detected with both instruments is shown in the left panel of Figure~\ref{f:970508} \citep{Piro98b}.  
About 6 h after the event, the~field obtained with \wfc\ was acquired by NFIs, and an unknown X-ray source (1SAX J0653.8+7916) was detected. Its coordinates were promptly distributed \citep{Costa97b}, an~optical source was discovered~\cite{Bond97}, and~soon later, several groups performed optical observations of the discovered source. The~detected optical counterpart showed a flux that was increasing after two days, arriving at a magnitude of $R = 20.14$ and then starting to fade with the power law (see~\cite{Piro98b} and references therein). 

On 11 May, when the optical afterglow was still relatively bright, our collaborators of the CalTech/NRAO group  observed it with the Keck Low
Resolution Imaging Spectrograph. Various absorption lines were identified (see right panel of Figure \ref{f:970508}): some at redshift $z= 0.835$, and some others at redshift of $z = 0.767$  \citep{Metzger97}. 
Weeks later, when the point-like object was almost invisible, the~highest redshift value ($z=0.835$) was found in the emission lines from an extended source in the same position of the point-like object: the galaxy that had hosted the fading object. 
The mystery of the GRB sites was solved.  Remote galaxies harbor~GRBs.

The immediate consequence of this discovery was that it was eventually possible to fix the energy scale.  From~the luminosity distance of the GRB~970508 optical counterpart ($1.49\times 10^{28}$~cm), obtained assuming a standard Friedmann cosmology with \mbox{$H_0 = 70$~km s$^{-1}$ Mpc$^{-1}$} and $\Omega_0 =0.2$, it was possible to derive the first estimate of the energy released by a GRB: $E_{iso} = (0.61\pm 0.13) \times 10^{52}$ ergs, assuming an isotropic~emission. 

GRB~970508 was also relevant for the discovery, with~the VLA radio telescope, of~the first radio afterglow \citep{Frail97}. The~radio emission from the optical GRB counterpart showed a phenomenon known as \textit{scintillation} that derives from the effects of interstellar clouds on sources of a very small angular size. In~GRB~970508, the~scintillation disappeared after about two months.
From the source angular size and its distance, it was possible to derive  the expansion velocity of the radio source \citep{Frail97}. It resulted to be around $2c$, an~apparent superluminal expansion, typical of sources expanding at relativistic~velocity.

%
%

On 14 December 1997, also the redshift of another \sax\ GRB (971214) was measured: z = 3.42. 
The corresponding energetics was $(2.45\pm 0.28)\times 10^{53}$ ergs corresponding to  0.14~M$_\odot$c$^2$, a~value never observed~before.

\section{Immediate Consequences of the \sax\ Discoveries}
\label{immediate consequences}

\subsection{Scientific Community~Reaction}
\label{reactions}

\begin{figure}[H]
    \includegraphics[width=.99\textwidth, angle=0]{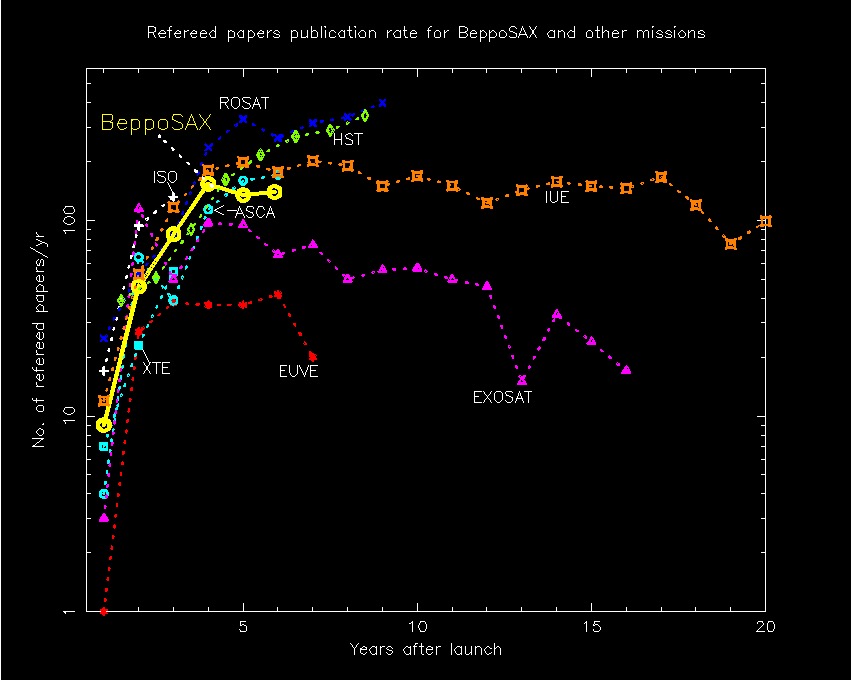}
\caption{{Number} 
 of citations of the papers based on the results with the  \sax\ observations (mainly GRBs), compared with those based on the results obtained with other satellites. Figure kindly received by Paolo Giommi, at~that time, the director of the ASI SDC (SAX Data Center).}
 \label{f:SAX-citations}
\end{figure}

\begin{itemize}
\item
In the first two years (1997--1998), the number of papers citing \sax\ was similar to those citing \hst\ (about 200/yr) (see Figure~\ref{f:SAX-citations}). The~Science/AAAS journal classifed GRB discoveries among the top ten over the world and over all the science fields.
\item 
The data flow from the \integral\ satellite was modified for a prompt localization of GRBs through an on-ground data analysis software of the IBIS instrument \citep{Mereghetti2000}.
\item
NASA issued an Announcement of Opportunity for a new medium-size satellite mission, that led to the \swift\ selection ({\em {\rm now} Neil Gehrels Swift Observatory}). The~\swift\ mission, still operational, has a configuration similar to  that of \sax, with~a wide field GRB monitor (\bat\ (Burst Alert Telescope), \citep{Barthelmy05a}) for the prompt identification and localization of GRBs, and~an X-ray Telescope (XRT, \citep{Burrows05}) plus an Ultraviolet/Optical Telescope (UVOT, \citep{Roming05}) for the afterglow observation. In~order to study the early afterglow, impossible with \sax\ (anything was known about the GRB evolution), the~GRB follow-up is automatically performed in a very short time ($\sim$100~s) \citep{Gehrels04}. 
\item
All scientists who managed large radio and optical telescopes devoted observation time to follow-up GRBs localized with \sax. Also the observation procedures and equipment were changed to make these observations faster.
\item 
Several new optical and/or NIR telescopes were built or modified to allow the robotic pointing of the \sax\ GRB events. 
\item
The \sax\ GRB coordinates were distributed through the already existing GCN (General Coordinates Network) circulars set up by NASA, which received an impressive boost by the \sax\ findings.
\item
Also, the \fermi\ high-energy gamma-ray satellite was designed taking into account the \sax\ payload configuration: a Gamma-Ray Burst Monitor {\em GBM} (8~keV--40~MeV over the full unocculted sky) to identify GRBs, and~a {\em LAT} gamma-ray telescope (20~MeV--300 GeV, 2.3~sr FOV) to localize and study them in the MeV/GeV energy range \citep{Atwood09,Meegan09}. 
\item
A similar design to \sax\ was adopted for the \agile\ Italian satellite, with~a hard X-ray imager (Super\agile) sensitive in the range 18--60 keV with about 1 sr FOV and a gamma-ray imager sensitive in the range 30~MeV--50~GeV \citep{Tavani08}. 
\end{itemize}

%
%

\subsection{Impact of the BeppoSAX Discoveries on GRB Theoretical~Models}
\label{s:theory}

\begin{figure}[H]
	\includegraphics[width=0.8\textwidth]{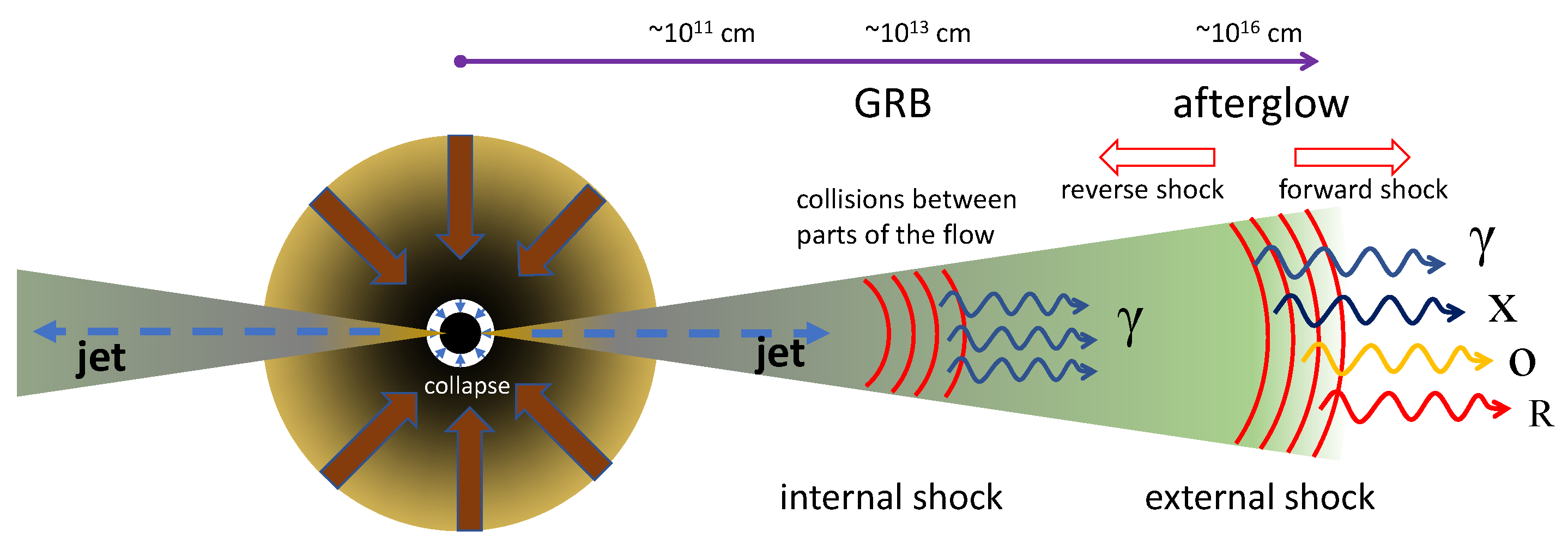}
		\caption[]{Sketch of a relativistically expanding fireball. The~kinetic energy of the jet  is converted through shocks (internal, external) in electromagnetic energy. Figure reprinted from~\cite{Dado22}. }
	\label{f:fireball}
\end{figure}

\textls[25]{The measured distance scale of \sax\ GRBs  swept away all the galactic models. 
The discovered properties of the events, like their huge isotropic energy (up to $\sim$$10^{54}$~erg), their non-thermal spectra, and their short time variability (down to ms time scale),  were generally interpreted as a result of the formation of a fireball in relativistic expansion (see Figure~\ref{f:fireball}). This model, already developed before the \sax\ discovery of the X-ray afterglow (e.g.,~\cite{Guilbert83,Goodman86,Paczynski86}), had immediate success for its capability to explain the spectral and temporal properties of the discovered GRBs (e.g., \cite{Wijers97,Sari98}), through the conversion of the fireball kinetic energy into electromagnetic radiation. This conversion was assumed to occur, for~the prompt emission,  through shocks between contiguous shells within the fireball, while, for~the afterglow emission, through shocks in the external medium (e.g., \cite{Meszaros94,Paczynski94}). }

However, the~smaller energy conversion efficiency in the internal shocks  than in the external shocks was noted to be inconsistent with the observation results (e.g., \cite{Daigne98}). Indeed, unlike the expectations from the fireball model, the~energy released during the prompt emission was higher than that in the afterglow, at~least on the basis of the afterglow spectrum, which was (and is still) possible to measure only up to 10~keV.

%
%

Concerning the GRB progenitors, for~short GRBs (T$_{90}< 2$ s), the merging of binary systems like  white dwarf (WD)--neutron star (NS)  or NS-NS or NS-BH, were considered the most likely mechanisms  \citep{Narayan92}. Instead, for~long GRBs (T$_{90}> 2$ s), failed supernovae~\cite{Woosley93} or the collapse to a Kerr black hole of a rapidly rotating star (collapsar) \citep{Paczynski98} with the formation of hypernovae, were the most favorite models. Other models were also suggested for the GRB production, like the supranova model \citep{Vietri98}, or~the transition, by~accretion,  of~a neutron star to a quark star~\cite{Berezhiani03}, or~the Electromagnetic Black Hole (EMBH) model \citep{Ruffini01}.

\section{Other Relevant Results Obtained with \sax, with~Most of Them Later~Confirmed}
\label{other sax discoveries}

Other relevant results concerning the prompt emission and the afterglow were obtained with \sax. Some of them, like the discovery of transient absorption lines and variable column density (see below), have still not been confirmed. This fact could be due to the passband of post-\sax\  wide field instruments, especially those that are more sensitive, whose lower energy threshold is in the hard X-ray band. The~\theseus\ mission concept \citep{Amati21} could definitely settle this issue. The~most relevant \sax\ results are discussed in the following~subsections.

\subsection{Discovery of Transient Absorption Lines in the Prompt~Emission}

\begin{figure}[H]
\includegraphics[width=0.4\textwidth]{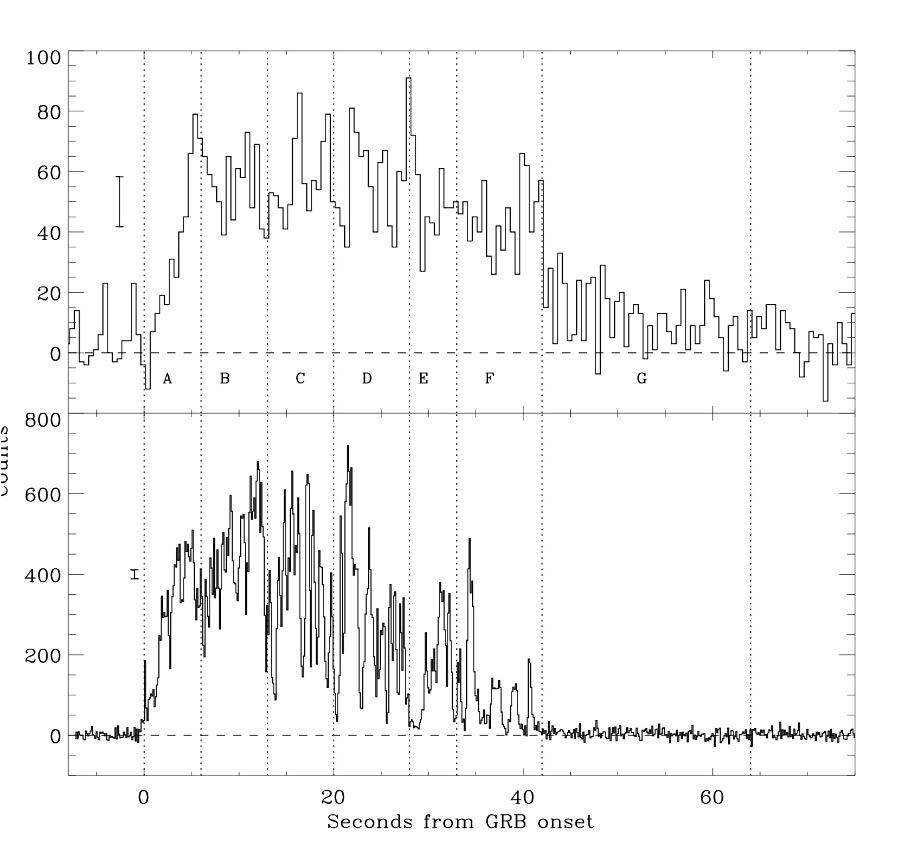}
\includegraphics[width=0.4\textwidth]{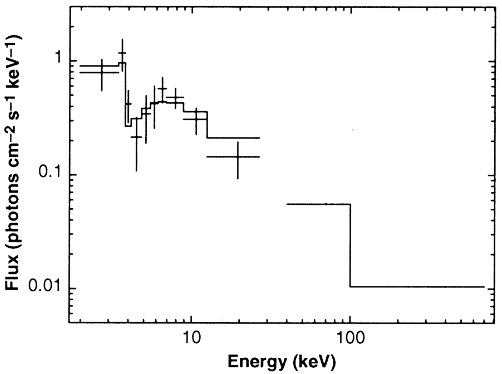}
 \caption[]{(\textbf{Top left}): {light} 
 {curve} 
 of GRB990705 obtained with \sax\ WFC (0.5 s time resolution). (\textbf{Bottom left}): light curve of GRB~990705 obtained with GRBM (0.128 s time resolution). The~vertical dotted lines limit the 7 intervals, named A, B, C, etc, in which the spectral analysis was performed. (\textbf{Right}): photon spectrum of GRB~990705 in the time slice B. The~best-fit curve is obtained with a power law plus a photoelectric absorption by a medium at redshift $z = 0.86$, column density N$_H = 1.3 \times 10^{22}$~cm$^{-2}$, and~iron abundance 75 times the solar one. A~similar fit was obtained for the spectrum of the time slice A, while the spectrum of the time slice C was well described by a simple power law. Figure reprinted from~\cite{Amati00}. }
\label{f:absorp-edge}
\end{figure}

\begin{itemize}
    \item 
A transient absorption edge at 3.8 keV in the prompt emission of the \sax\ GRB~990705  was first discovered \citep{Amati00}. The~results are shown in Figure~\ref{f:absorp-edge}. The~feature was found to be  consistent with a red-shifted K-edge due to an iron environment. The~confidence in the reality of this line is that the derived redshift ($z = 0.86$) was later measured from the GRB host galaxy \citep{Lefloch02}.
\item 
Investigating the spectral evolution of the prompt emission from the \sax\ GRB~011211 with measured redshift ({$z$} 
 = 2.140),  also evidence of a transient absorption feature at  $6.9^{+0.6}_{-0.5}$~keV during the rise of the primary event \citep{Frontera04} was found. The~significance of the feature was derived with non-parametric tests and numerical simulations, finding a chance probability of $3 \times 10^{-3}$ down to $4 \times 10^{-4}$. The~feature showed a Gaussian profile and an equivalent width of about 1~keV. See~\cite{Frontera04}, where a possible interpretation is also discussed.
\end{itemize}
%
%
%
\unskip
  
\subsection{Detection of a Transient Column Density in the Prompt~Emission}

\begin{figure}[H]
\includegraphics[width=0.5\textwidth]{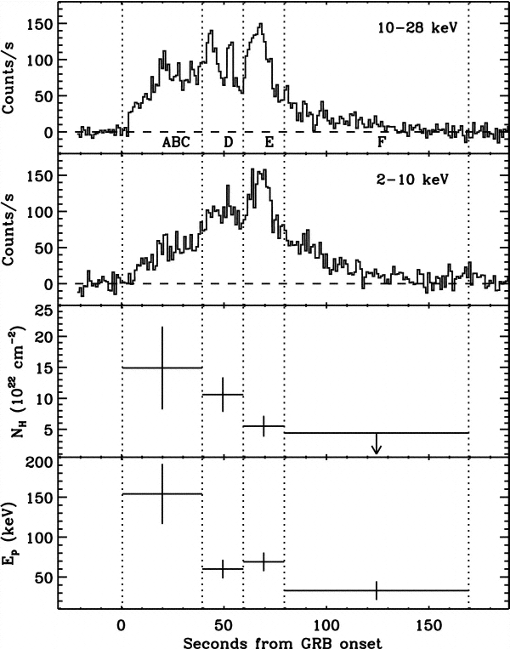}
 \caption[]{Measured column density as function of time from \sax\ GRB~000528 onset. Figure reprinted from~\cite{Frontera04a}. }
	\label{f:decreasingNH}
\end{figure}

With {\sax}, thanks to the \wfcs\ low energy threshold, also the detection of a decreasing column density during the prompt emission of GRB~000528 \citep{Frontera04a}  was found (see Figure~\ref{f:decreasingNH}). This result is clear evidence that the GRB environment was ionized by the gamma-ray~event.

%
%

Unfortunately, given the higher value of the low energy threshold of the later GRB monitors (at a minimum, 8 keV with \swift\ \bat\ and \fermi\ \gbm), similar searches could not be performed until now. In an~investigation  performed using the \swift\ XRT data of 199~GRBs, ~7 GRBs, in~late time intervals (between about 1 and 2 min from the GRB onset), showed signs of a decrease in N$_H$ (see review paper \citep{Valan23}). 
A definitive response to a ionization of the GRB environment could be given by the launch of the \theseus\ mission concept (see~Section {\ref{s:future missions}).} 

\subsection{Discovery of the GRB--Supernova~Connection} 

For the first time, a~\sax\ GRB (980425) was found to be the likely origin of a type Ic supernova (SN) explosion: SN1998bw \citep{Galama98b}. The~direction of this supernova had resulted to be coincident with that of GRB~980425, and~its explosion was contemporary, within~one day, with~the occurrence of the  GRB event. The~SN was unusually bright (for this reason, it was called hypernova) and was expanding at a very high velocity \citep{Patat01}.  

The uncertainty about the chance coincidence of SN1998bw with GRB~980425 was definitively removed in 2003, when the type Ic SN2003dh was found to be associated with GRB\,030329 \citep{Stanek03}. Indeed, in~this case, the~spectra of the optical emission, initially consistent with a power-law continuum, after~a week, became remarkably similar to those of SN1998bw, once they were corrected for the afterglow~emission.  

Nowadays, it is a matter of fact that several long GRBs originate in supernova explosions. In~Table~\ref{t:SNe-GRBs}, we list the events with identified SNe, obtained by updating the list given by~\cite{Cano17}. In~addition to these identified SNs, many other GRBs have optical counterparts with an afterglow curve that exhibits a clear bump plus spectroscopic signatures typical of a SN (see review by~\cite{Cano17}). 

\begin{table}[H]
\caption{{List of} 
 GRBs associated with identified supernovae (SNe). ll $=$ low-luminosity GRB ($L_{\gamma,iso} < 10^{48.5}$ erg~s$^{-1}$), INT $=$ intermediate-luminosity GRB ($10^{48.5} < {\gamma,iso}<10^{49.5}$ erg~s$^{-1}$). \mbox{UL $=$ ultra-long GRBs ($\sim 10^4$~s).}}
\label{t:SNe-GRBs}
	
		\newcolumntype{K}{>{\arraybackslash}X}
		\begin{tabularx}{\textwidth}{CCCC}
			\toprule
			\textbf{GRB}	& \textbf{Redshift, $z$}	& \textbf{GRB Type}     & \textbf{SN Search}\\
			\midrule
			980425		& 0.0085		& long, ll	   & 1998bw\\
			011121	    & 0.362			& long		    & 2001ke \\
                021211      & 1.004         & long          & 2002lt \\
                030329      & 0.16867       & long          & 2003dh \\
                031203      & 0.10536       & long, ll      & 2003nw \\
                050525A     & 0.606         & long          & 2005nc \\
                060218      & 0.03342       & long, ll      & 2006aj \\
                081007      & 0.5295        & long          & 2008hw \\
                091127      & 0.49044       & long          & 2009nz \\
                100316D     & 0.0592        & long, ll      & 2010bh \\
                101219B     & 0.55185       & long          & 2010ma \\
                111209A     & 0.67702       & UL            & 2011kl \\
                120422A     & 0.28253       & long          & 2012bz \\
                120714B     & 0.3984        & long, INT     & 2012eb \\
                130215A     & 0.597         & long          & 2013ez \\
                130427A     & 0.3399        & long          & 2013cq \\
                130702A     & 0.145         & long, INT     & 2013dx \\
                130831A     & 0.479         & long          & 2013fu \\
                161219B     & 0.1475        & long, INT     & 2016jca \\
                171010A     & 0.3285        & long          & 2017htp \\
                171205A     & 0.037         & long, ll      & 2017iuk \\
                180728A     & 0.117         & long          & 2018fip \\
                190114C     & 0.4245        & long          & 2019jrj \\
                190829A     & 0.07          & long, ll      & 2019oyw \\
			\bottomrule
		\end{tabularx}

\end{table}

In general, the associated SNe are type Ic with high expansion velocities and much larger energy release  than  in  normal SNe, and thus are hypernovae as found in the case of SN1998bw.  In~addition, as can be seen from Table \ref{t:SNe-GRBs}, most of them have a known redshift with values $<1$.

However, there are GRBs with~the same features of events associated with a SN (like their long duration, low luminosity, and redshift $<1$) with  no associated SN (see~\cite{DellaValle06a}), demonstrating that there are GRBs originating in very faint supernovae or they are due to different~phenomena. 

%
%

%

\subsection{Discovery of the $E_p$--$E_{iso}$ Relation}

\begin{figure}[H]
	\includegraphics[width=0.8\textwidth]{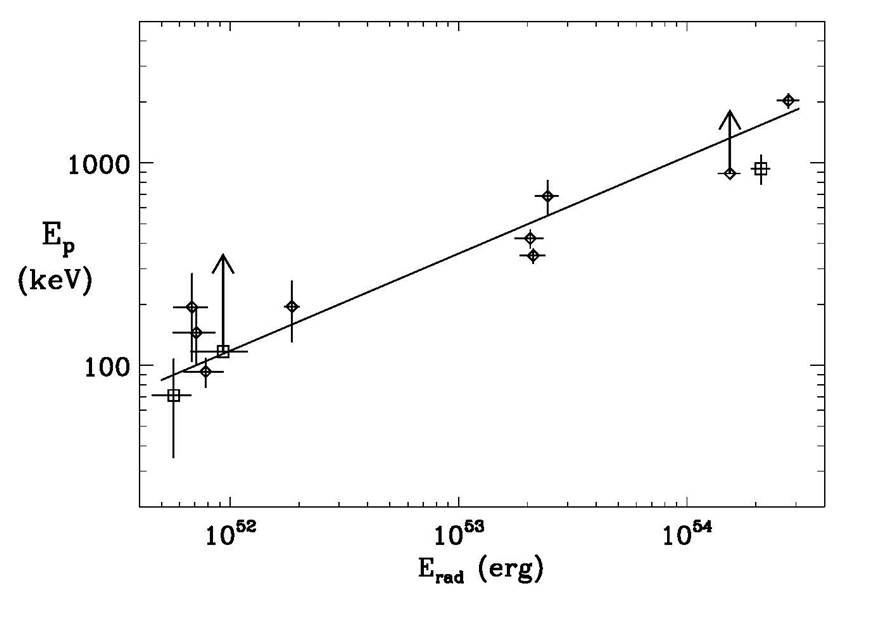}
		\caption[]{The $E_{p,i}$--$E_{iso}$ relation discovered with \sax. Reprinted from~\cite{Amati02}.}
	\label{f:ep-eiso}
\end{figure}

This relation, now also known as the {\em Amati relation} from the name of the first author of the discovery paper \citep{Amati02}, gives the correlation between the photon energy, redshift-corrected, at~the peak of the prompt spectrum $\nu F_\nu$ of a long GRB and the total energy released during the burst $E_{iso}$, assuming isotropy in the emission (see Figure~\ref{f:ep-eiso}). This relation was discovered with a set of 12 \sax\ long GRBs ($>$2~s), whose redshift was determined with optical spectrometers \citep{Amati02}. Two years later, after~we published this result, two other relationships were reported: the ``Yonetoku relation'' between rest frame $E_{peak}$ and bolometric peak luminosity $L_{p,iso}$ \citep{Yonetoku04}, and~the ``Ghirlanda relation'' between rest frame $E_{peak}$ and released energy $E_\gamma$ after correction for the beaming factor ($E_\gamma = (1- cos\theta)E_{iso}$, where $\theta$ is the angular width of the jet), assuming a jet-like emission \citep{Ghirlanda04a}. For~a discussion on the weak points of the Ghirlanda relation, see~\cite{Frontera12a}. 

\vspace{-6pt}
%
%

Also, relations between prompt and afterglow emission or about only the afterglow were reported. For~a review, see \citep{Dainotti17}. 

 Among these relations, the~$E_{p}$--$E_{iso}$  relation remains the most robust. It is now confirmed that this relation is satisfied by almost all long GRBs, for which, along with $z$, it has been possible to estimate released energy and peak energy $E_p$. The~only outliers are two long GRBs: {980425} 
 associated with SN1998bw, and~031203 associated with SN2003nw (see Table~\ref{t:SNe-GRBs}).
Possible explanations for these apparent outliers have been investigated. According to~\cite{Ghisellini06,Ghirlanda07c}, from~the fact that these two bursts share several properties with GRB~060218, an~event associated with SN2006aj, which, however, obeys the $E_p$--$E_{iso}$ correlation, these authors suggest that such discrepancy could be due to the limited energy band in which these two events have been observed that could have biased the derived $E_p$ value.

  Some authors suspected that the  $E_{p}$--$E_{iso}$ relation could be influenced by selection effects (see \cite{Butler09}). However,  by~slicing the GRB time profiles in several time intervals and deriving the spectra in each of them, the~correlation between the time resolved  $E_p$ and the corresponding flux is still apparent (see~\cite{Ghirlanda10,Frontera12a,Maistrello24}).  A~possible origin of the correlation was also discussed by~\cite{Frontera16}. 

\begin{figure}[H]
	\includegraphics[width=0.6\textwidth]{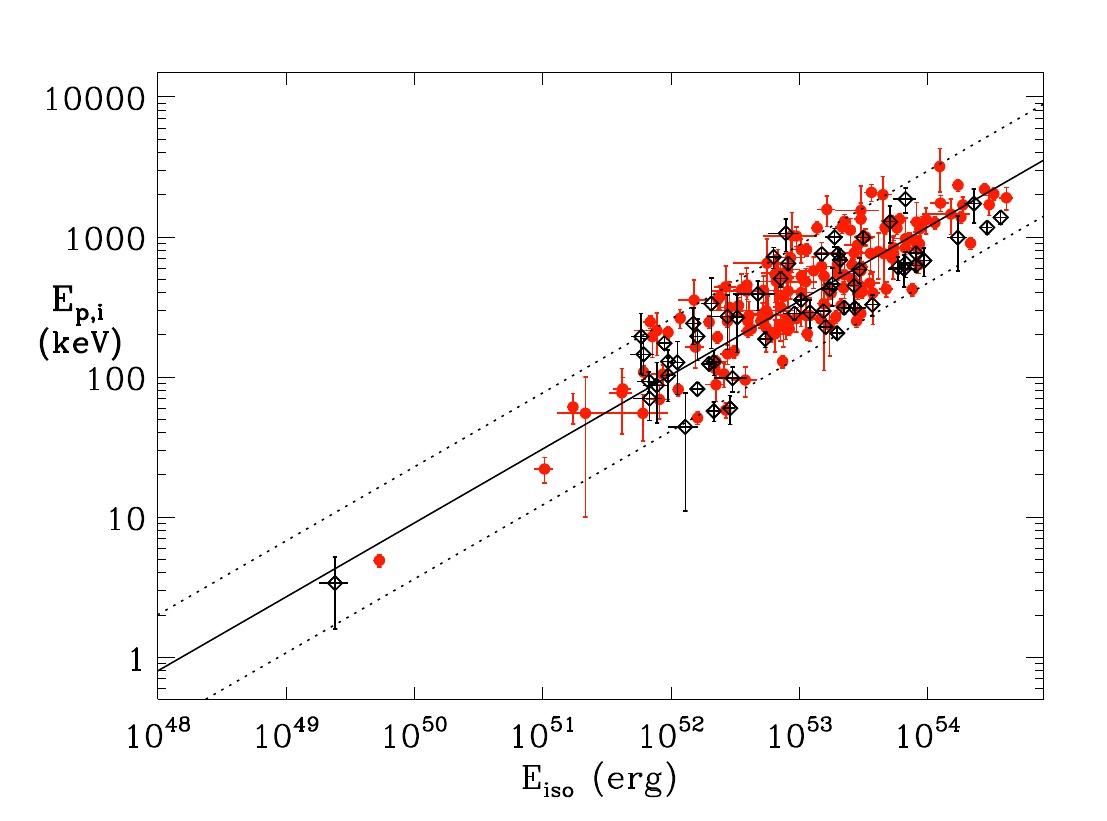}
		\caption[]{{The} 
 $E_{p,i}$--$E_{iso}$ {correlation} 
 {for} 
 a recent sample of 224 long GRBs. Those events detected and localized by the \swift\ satellite are shown in red color, while those in black color were detected with other satellites (\sax, \integral, \fermi, etc). Courtesy of Lorenzo Amati.}
	\label{f:newep-eiso}
\end{figure}

 The Amati relation also appears to be a promising tool to describe the expansion history of the universe and to estimate the cosmological parameters (see~\cite{Demianski17,Amati13}). A~recently updated $E_{p}$--$E_{iso}$ relation is shown in Figure~\ref{f:newep-eiso}. The~relation is discussed in the context of a review on cosmological probes \citep{Moresco22}.

\vspace{-12pt}
%
%

\subsection{Discovery of X-ray Flash and X-ray Rich~Events} 

\subsubsection{X-ray Flash~Events}

X-ray flashes (XRFs) were discovered with \sax\ as bright, low-energy events with a duration like that of long GRBs. They were detected with the WFCs (2--28 keV) but undetected by GRBM (40--700 keV) \citep{Heise01}. Their temporal and spectral properties were found to be very similar to those of the X-ray counterparts of GRBs \citep{Heise03}. 

These events were later better investigated with the \hete\ satellite \citep{Lamb04a}) launched in October 2000, and~then with the \swift\ satellite (see below). Thanks to \hete\, it was possible to establish (see~\cite{Pelangeon08}) that X-ray flashes show properties similar to those of long GRBs but~with a lower peak energy $E_p$ (see top panel of Figure~\ref{f:xrayflashes}) and a lower redshift distribution (see bottom panel of Figure~\ref{f:xrayflashes}).

%
%

\vspace{-6pt}
\begin{figure}[H]
	\includegraphics[width=0.75\textwidth]{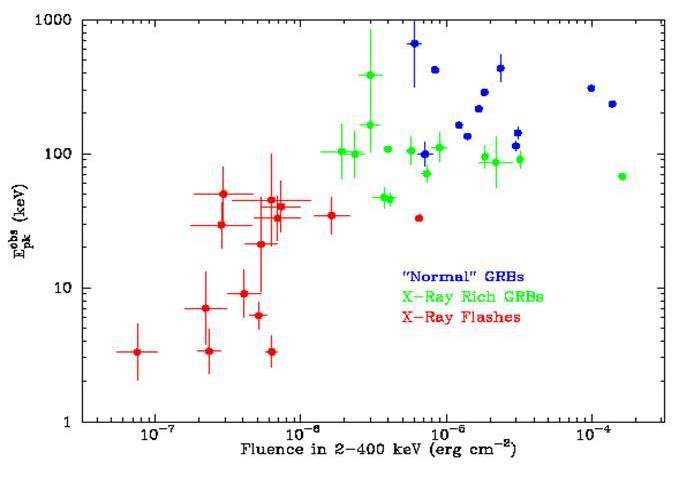}
			\caption[]{\emph{Cont}.} 
\label{f:xrayflashes}
\end{figure}

	\begin{figure}[H]\ContinuedFloat
	\includegraphics[width=0.75\textwidth]{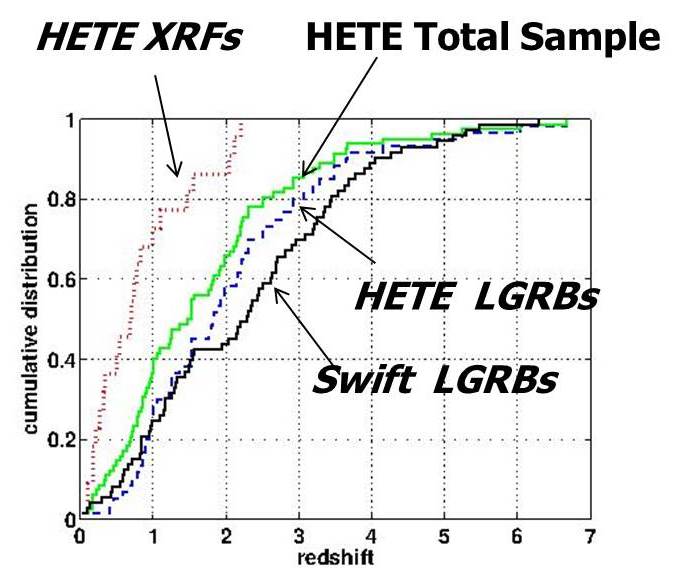}
		\caption[]{(\textbf{Top panel}): Distribution of the peak energy $E_p$ vs. fluence for XRFs, X-ray rich (XRR) events, and~GRBs. Adapted from~\cite{Sakamoto05}, where you can find a discussion on XRF and XRR events.  (\textbf{Bottom panel}): {redshift} distribution of GRBs and XRFs. Figure adapted from that of~\cite{Pelangeon08}.} 
\label{f:xrayflashes}
\end{figure}

XRFs were further investigated with \swift. They were found to have a $T_{90}$ duration between 10 and 200 s and an isotropic sky distribution. Thus, in~these respects, XRFs are similar to “classical” GRBs. Also, the spectral analysis showed that their spectra were similar to those of long GRBs, with~the main difference that XRFs have peak energies $E_p$ of their $EF(E)$ spectra that are much lower than those of long GRBs. Therefore, XRFs could be a subpopulation of GRBs with low peak~energies.

An interesting discovery, in~favor of the GRB-like picture, was the observation of an XRF event (060218) associated with the type Ic SN2006aj supernova \citep{Pian06}.
SN2006aj was intrinsically less luminous than other previously discovered SNe associated with GRBs but~more luminous than many supernovae not accompanied by a GRB. From~the weakness of both the GRB output and the supernova radio flux, it could be established that XRF~060218 was an intrinsically weak and soft event rather than a classical GRB observed off-axis. These results extended the GRB--supernova connection to X-ray flashes and fainter supernovae, implying a common~origin.

\subsubsection{X-ray Rich~Events}

Also, X-ray rich GRBs (XRRs) were discovered with \sax\ \citep{Heise03}.  Their properties were found to be intermediate between XRFs and classical~GRBs.  

The XRF and XRR interpretation as a subclass of classical GRBs  is now strongly confirmed: see a recent review on the XRF properties by~\cite{XiongweiBu18}, in~which it is definitely found that both XRF and XRR events are GRBs  with the following main~properties:
\begin{itemize}
\item [(a)] Low values of $E_p$, thus main fraction of energy released in the X-ray band; 
\item[(b)] Low luminosity; 
\item[(c)] Long duration; 
\item[(d)] $E_{p}$--$E_{iso}$ correlation as GRBs; 
\item[(e)] Same redshift distributions as GRBs; 
\item[(f)] The association with SN explosions is favored.
\end{itemize}

\subsection{Spectral Properties of the Late Afterglow of Long~GRBs}
\label{s:spectral-properties}

Concerning the late GRB afterglow, \sax\ also allowed to derive its average spectral  properties with the LECS/MECS telescopes (0.2--10 keV). The~average properties of the late afterglows were summarized as follows \citep{Frontera03}:
\begin{itemize}
\item
Photon spectra consistent with a photo-electrically absorbed power law, with~photon indices $\alpha$ distributed according to a Gaussian with average photon index\linebreak \mbox{$\alpha_{ave} = 1.95 \pm 0.03$} and standard deviation $\sigma_{\alpha} = 0.4$.
\item
Fading behavior also consistent with a power law with index $\beta$ distributed according to a Gaussian ($\beta_{ave} = -1.30 \pm 0.02$; $\sigma_{\beta} = 0.35$).
\end{itemize}

With \swift, the~late afterglow spectral properties have been fully confirmed, while the temporal properties, extended to the early afterglow, have been found to show a much more complex behavior (see Section~\ref{s:swift-results}).

About the hard X-ray band, with~\sax\, it was also detected the first X-ray afterglow above 10 keV. This detection was obtained in the case of  the very strong GRB~990123. In~this case, it was possible to detect the afterglow spectrum up to 60 keV with the PDS instrument~\citep{Maiorano05}. The~high-energy spectrum was found not to be consistent with the power-law spectrum measured below 10 keV with LECS/MECS. An~interpretation of the hard X-ray spectrum in terms of an Inverse Compton (IC) component in addition to a synchrotron spectral emission was also discussed \citep{Corsi05}.

\textls[-15]{After \sax, the~hard X-ray afterglow spectrum was detected in only two cases of very bright GRBs. One case is the long GRB~120711A detected with the \integral\ satellite~\citep{Martin-Carrillo14} for 10 ks in 20--40 keV, and, for~a shorter time, at~higher energies. According to these authors, likely the long duration of the high energy emission was still prompt emission. In~any case, for~this GRB, by~combining the \integral\ IBIS and the \fermi\ LAT instruments, the~hard X-ray energy spectrum was found consistent with synchrotron~radiation. }

\textls[-15]{Another case was that of the bright GRB~130427A observed with \nustar\ \mbox{(3--79 keV)}~\citep{Kouveliotou13}, starting approximately 1.2, 4.8, and~5.4 days after the \swift\ GBM trigger. In~this case, it was shown that the hard X-ray observations (up to 80 keV) were crucial to establish that the afterglow was due to synchrotron radiation, and~to provide a strong direct observational support for such an emission mechanism also in the high-energy gamma-ray band, investigated with the \fermi\ mission. }

Hard X-ray afterglow emission up to about 20 keV was also measured  with \nustar\ in the case of GRB~130925A \citep{Bellm14}, but given the narrow band of the GRB detection in hard X-rays, no definite conclusion about the origin of the hard X-ray component could be~drawn.

\subsection{BeppoSAX GRBM Catalog of~GRBs}

\begin{figure}[H]
	\includegraphics[width=0.85\textwidth]{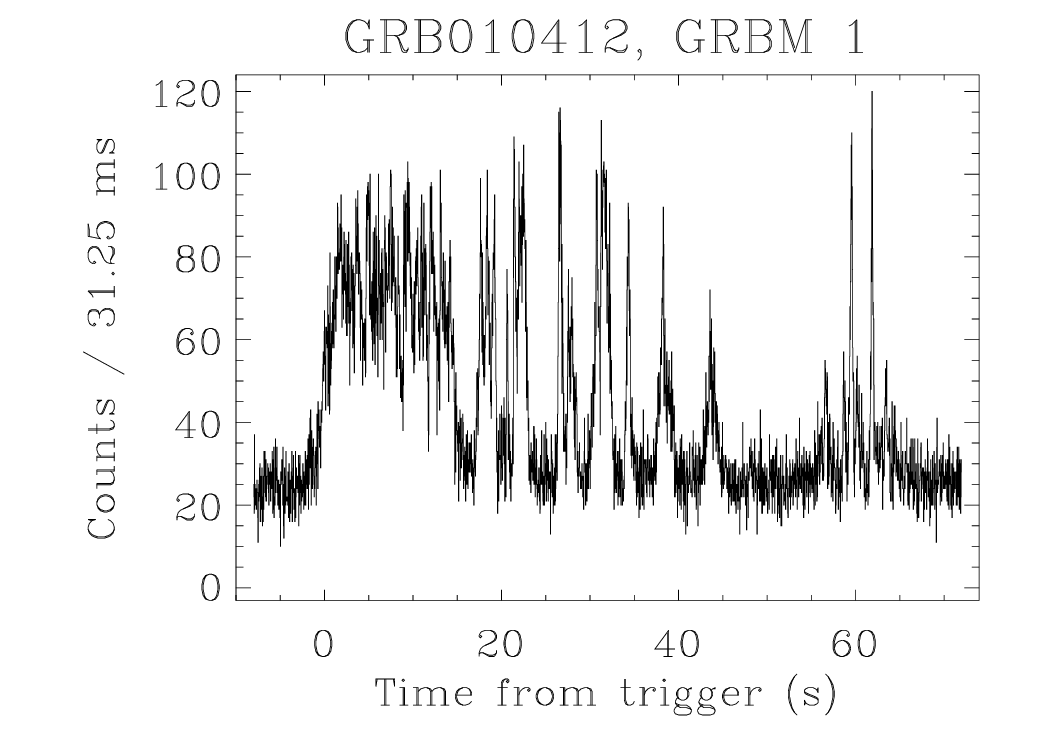}
	\vspace{-3pt}
		\caption[]{GRB~010412 as detected by GRBM aboard \sax, with~7.8 ms time resolution.}
\label{f:grb010412}
\end{figure}

The \sax\ GRBM instrument detected  1082 GRBs (see catalog in \citep{Frontera09}) with \mbox{40–700 keV} fluences in the range from $1.3 \times 10^{-7}$ to $4.5 \times 10^{-4}$~erg~cm$^{-2}$, with~the discovery of very peculiar events, like that shown in Figure~\ref{f:grb010412},  featuring many peaks with subsecond duration and the finding that the GRB active time (i.e., the~time $T_a$, in~which the burst was visible above a 2$\sigma$ level) has a cutoff at about 200~s, in~spite of the fact that the found GRBs had a total duration up to 600 s. This result could be related to the fact that the inner engine could have a maximum active time of about 200~s.

Also, a spectral catalog was derived from the GRB detections \citep{Guidorzi11}, in~which the average spectra of the 200 brightest GRBs were reported and are publicly available. Most of the photon spectra were found to be consistent with a  Band function, with~low-energy index around 1.0 and a peak energy $E_p$ of the $EF(E)$ spectrum of about 240~keV, in~agreement with previous results obtained on bright GRBs with the {\em CGRO} BATSE experiment. It was interesting to see the fact that about 30\% of the GRB spectra could be fit with a simple power law, suggesting that the peak energy could be close to or outside the GRBM energy boundaries as~later extensively confirmed with broader passband missions, e.g.,~\fermi.
%
%
\unskip

\section{New Discoveries on GRBs in the Post-\sax\ Era}
\label{post-sax era}

In addition to the confirmation of the huge discoveries obtained with {\sax}, many questions left unanswered by this satellite were solved in the post-\sax\ era, like the early afterglow properties, the~breaks in the X-ray afterglow light curves, the~afterglow of short bursts and their origin, and~the GRB~environment.

\subsection{Swift and Its Role in the Progress of the GRB Phenomenon}
\label{s:swift-results}

As already mentioned, the~\swift\ satellite \citep{Gehrels04}, launched on 20 November 2004 and still operational at this time, is one of the post-\sax\ missions that has already given and is still giving a very high contribution to the GRB astrophysics. The~payload configuration is similar to that of \sax\, with~a hard X-ray (15--150 keV)  wide field (1.5 sr) BAT telescope for GRB detection and localization, and~a focusing XRT telescope (0.1--10 keV) for the afterglow measurement and study. Thanks to its autonomous slew decision capability, \swift\ can start the afterglow observations in less than 2 min time. A~further \swift\ performance is due to a UV--Optical Telescope (UVOT) on board, that allows to identify the optical counterparts of the X-ray afterglow sources with a limiting magnitude, in~1000~s, of~\mbox{$B = $ 22.3} in white~light.

Thanks to this configuration, its rapid reaction capability and its long life time (already about 20 years), several new results have already been obtained with this mission, with~reviews of the most important ones, inclusive of those obtained from statistical studies, already reported (see~\cite{Gehrels09,Kumar15,Gehrels17,Le17,Li18,Li21,Li22,Deng22,Oates23,Troja23}).
Among them, two important results, left open by \sax\ and solved with \swift\ thanks to its rapid reaction capability,  are the early afterglow properties of long GRBs and the afterglow of short GRBs (no afterglow of short GRBs was discovered with \sax).

\begin{figure}[H]
	\includegraphics[width=0.9\textwidth,clip]{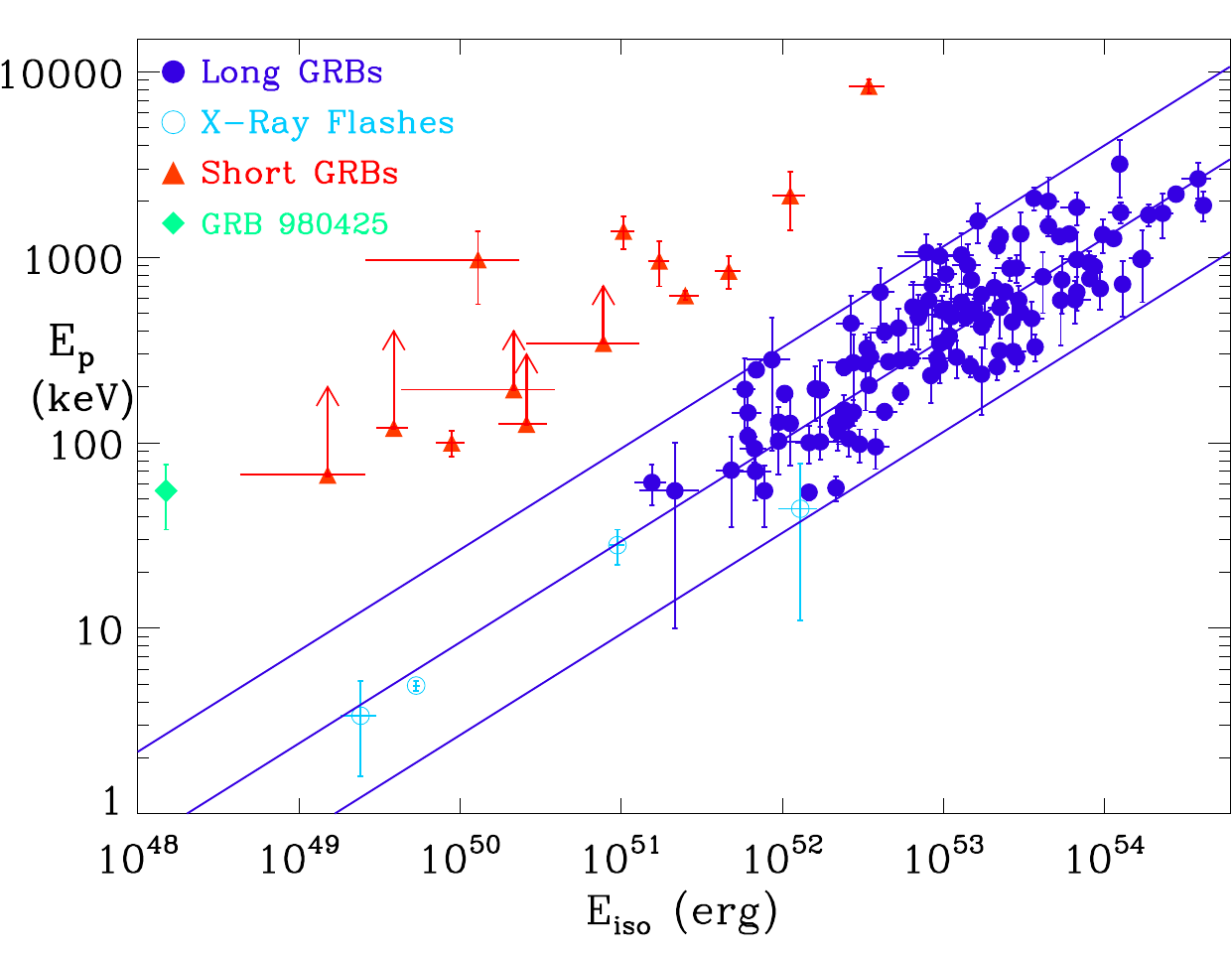}
		\caption[]{{The} 
 $E_p$--$E_{iso}$ relation in which short GRBs are also reported. As~can be seen, short GRBs are outliers of the Amati relation but~form a similar relation. Courtesy of Lorenzo Amati.}
	\label{f:shortgrb-EpEiso}
\end{figure}

The afterglow detection of short GRBs allowed  to determine their optical counterparts and thus their distances with the consequent determination of the energy released in these events, their redshift distribution and other properties, among~them being the~discovery that short GRBs are outliers of the Amati relation (see~Figure~\ref{f:shortgrb-EpEiso}), forming a similar relation offset from the main correlation.
%
%

\vspace{-6pt}

\begin{figure}[H]
	\includegraphics[width=0.8\textwidth]{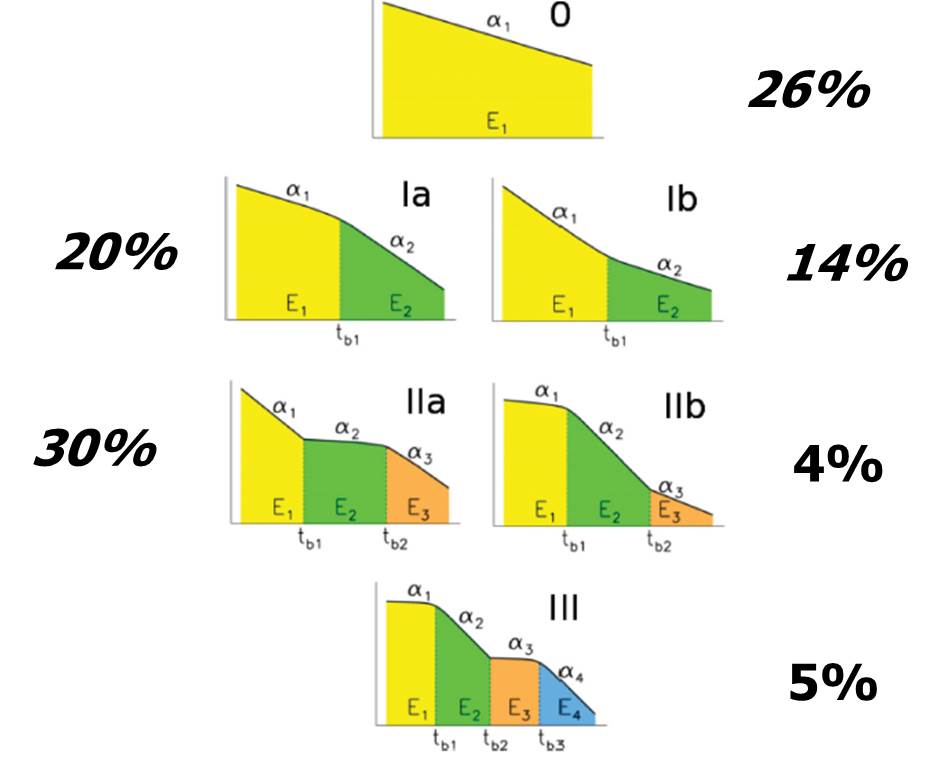}
		\caption[]{{Different} 
 types of afterglow light curves in 0.1--10 keV observed with \swift\ starting from \mbox{t $\gtrsim 60$~s.} Also, the fraction of GRBs with the given type  is shown. Initially, it seemed that the canonical light curves were either the IIa, whose frequency is only 30\% of the times, or~the Ia, whose frequency is 20\%. Figure adapted from that of~\cite{Margutti13}.}
	\label{f:early-afterglow}
\end{figure}

\begin{figure}[H]
	\includegraphics[width=0.8\textwidth]{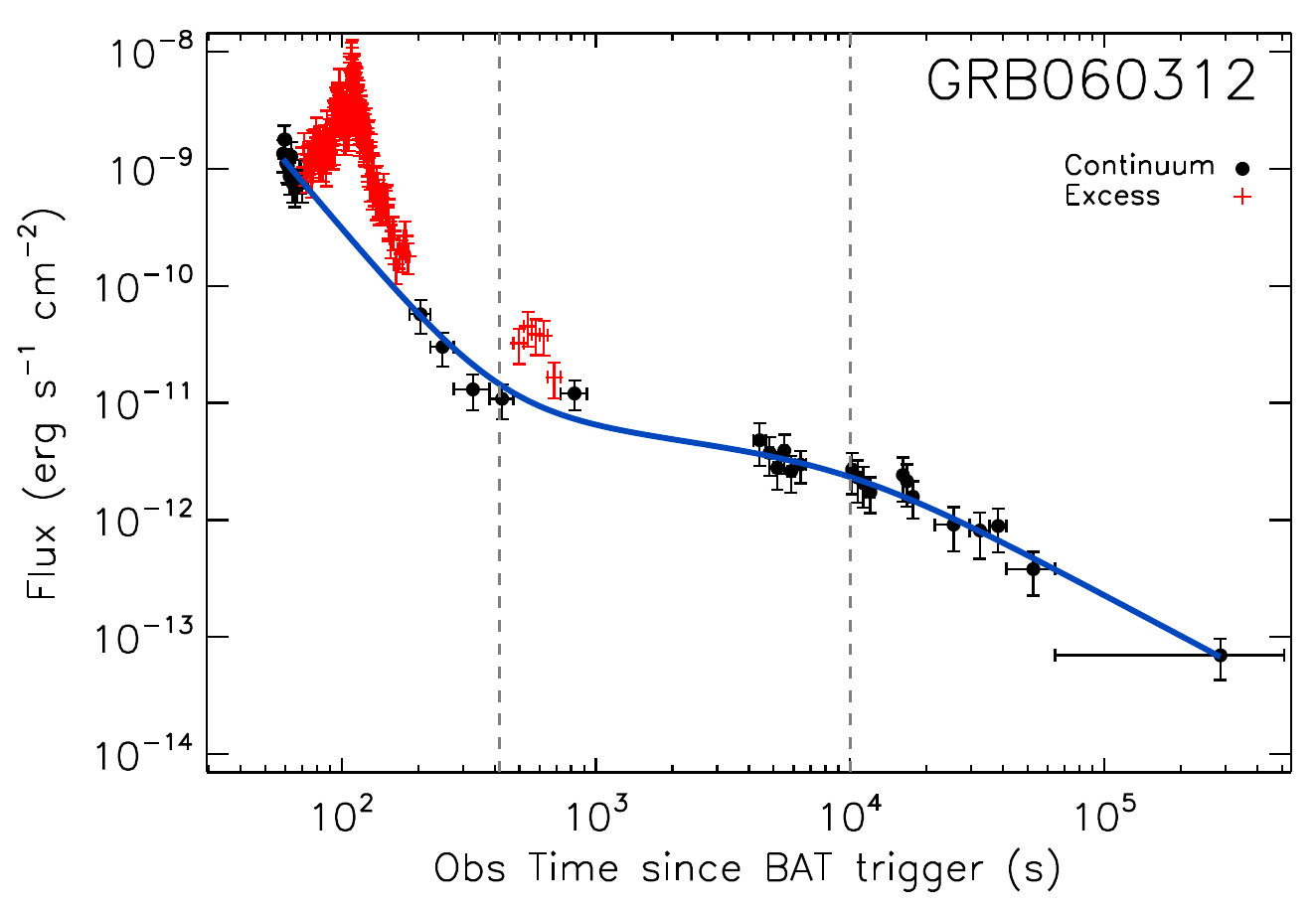}
		\caption[]{Example of light curve  of the intrinsic 0.6--30 keV afterglow starting from $T \ge 60$~s as~detected with \swift. Two X-ray flares above the smoothed light curve are apparent. Reprinted from~\cite{Margutti13}.}
	\label{f:060312-afterglow}
\end{figure}

Concerning the afterglow behavior from early to late times, a~first study based on 27 long GRBs \citep{Nousek06} observed with \swift~showed the presence of a canonical light curve, with~three distinct power-law decays ($\propto t^{-\beta}$): initially ($\lessapprox 500$~s) very steep ($ 3 \lessapprox \beta \lessapprox 5$), then ($ 10^3 \lessapprox t \lessapprox 10^4$~s) very shallow ($0.5 \lessapprox \beta \lessapprox 1$), and~finally, after~several hours, a~decay with $ 1.0 \lessapprox \beta \lessapprox 1.5$, as~found with \sax.
With a much more significant sample of \swift\ GRBs (622 long, and~36 short), the~afterglow properties were deeply investigated by~\cite{Margutti13}, inclusive of a statistical analysis of a possible relation of the X-ray afterglow properties with the prompt $\gamma$-ray emission. Their results showed that, unlike what was initially found~\citep{Nousek06}, the~light curves exhibit a variety  of power-law slopes (see Figure~\ref{f:early-afterglow}), with, in~some cases,  X-ray flares during the early afterglow, superimposed to smoothed power laws (see~Figure~\ref{f:060312-afterglow}).

\begin{figure}[H]
\includegraphics[width=0.8\textwidth]{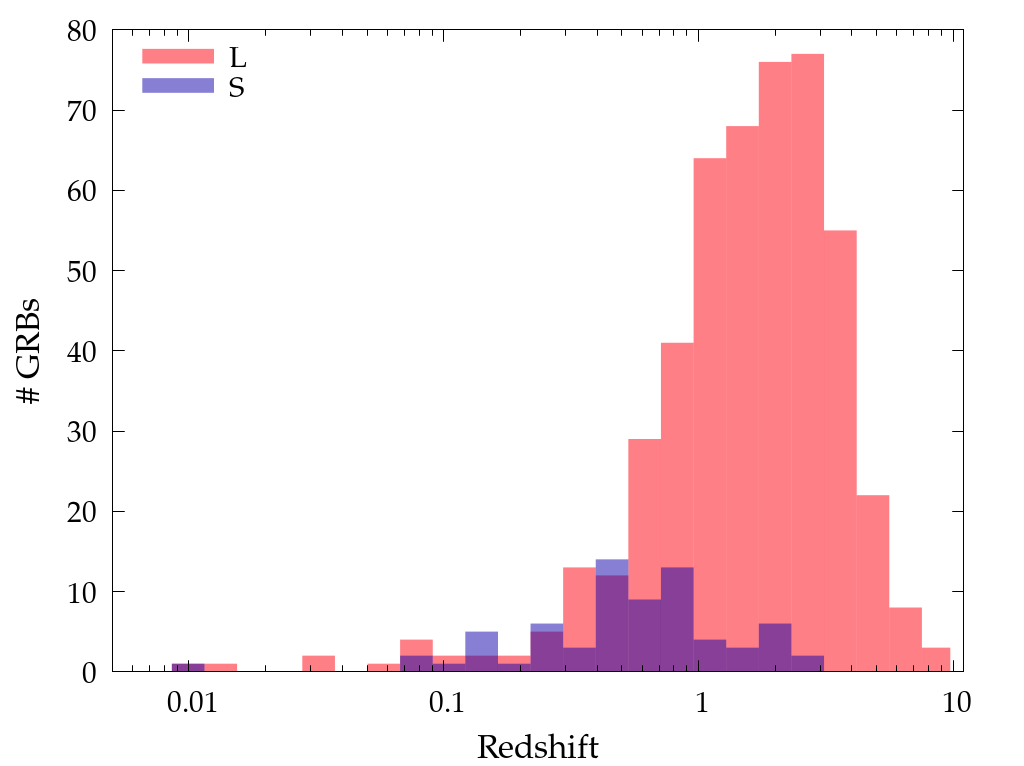}	
 \caption[]{Redshift distribution of 70 short (purple) and 488 long GRBs, list updated up to April 2024. Most of them have been discovered and promptly localized with \swift. The smaller distance scale of short GRBs is clear. Figure kindly provided by Cristiano Guidorzi.}
	\label{f:z-distribution}
\end{figure}

Among the many other results, thanks to the \swift\ configuration and its prompt dissemination of the GRB locations, it merits mentioning the possibility allowed by \swift\ to perform a prompt optical and radio follow-up, with~the determination of the redshift of a significant number of optical/radio GRB counterparts (see Figure~\ref{f:z-distribution}). As~can be seen from the figure, the~redshift distribution extends up to high $z$, with~the highest redshift events being GRB~090423 with $z = 8.26$ and GRB~090429B with $z = 9.4$.

Studies of the GRB distribution with $z$ (see~\cite{Robertson12}) have shown a significant correlation between the density of the GRB rate, with~redshift between 0 and 4, and~ star formation rate (SFR) density in the same redshift interval. The~ratio $\Psi (z)$  between the GRB rate and SFR is found to follow a power law ($\Psi (z)\propto (1+z)^{0.5})$ which can be explained if GRBs occur in low-metallicity galaxies. If~this relation would continue for $z>4$, the~discovery rate of very distant GRBs would imply a SFR density much higher than that inferred from UV-selected galaxies. The~discovery of a significant sample of high $z$ GRBs is still not obtained and is expected to be obtained, e.g.,~by the mission concept \theseus.

%
%
%
%
%

%

\vspace{-12pt}

\vspace{-6pt}
%
\subsection{The Contribution to GRB Studies by High-Energy Gamma-Ray Space Missions and  VHE Ground~Telescopes}

\begin{figure}[H]
	\includegraphics[width=1.0\textwidth]{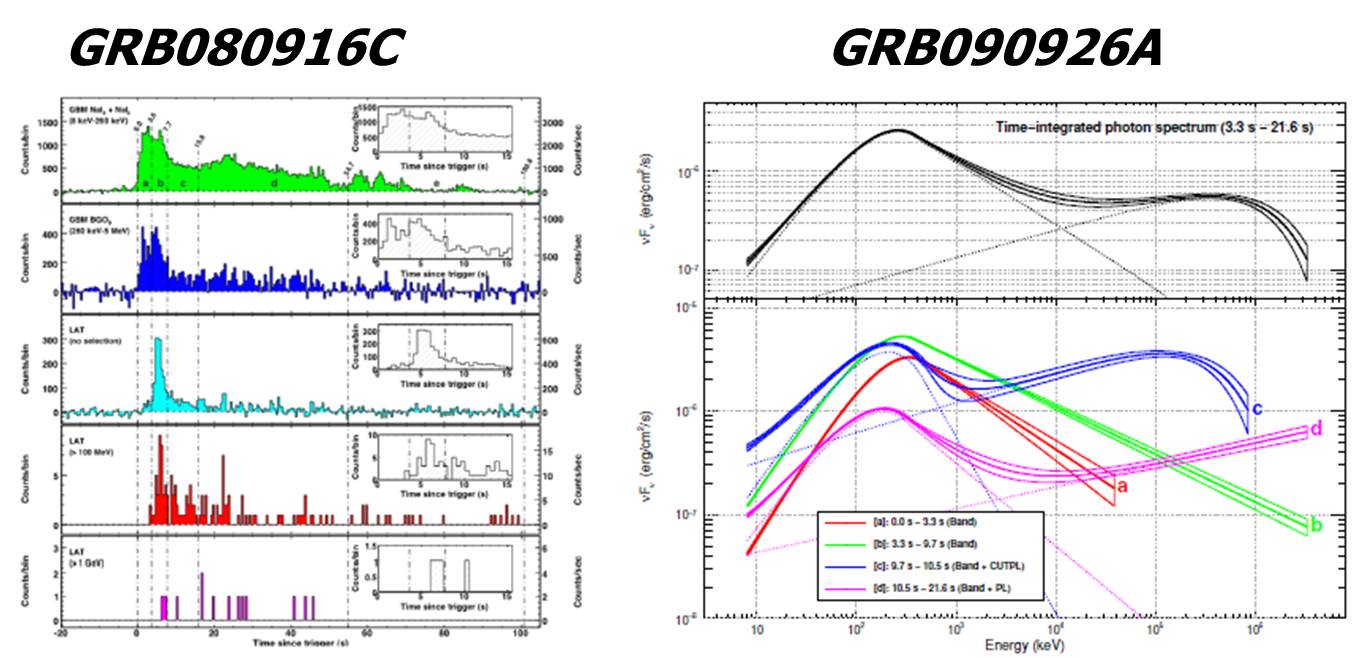}
		\caption[]{ (\textbf{Left panel}): {Light } 
{curve} 
 of the \fermi\ GRB~080916C at different energies. The delay of the onset of the high energy gamma-rays is apparent with respect to the onset of the GRB at low energies. Adapted from~\cite{Abdo09a}. 
  (\textbf{Right panel}): Time-resolved spectra of the \fermi\ GRB~090926A. A~hardening of the high-energy spectrum with time is apparent when the high-energy gamma-ray spectral component appears during the latest time interval. Reprinted from~\cite{Ackermann11} © AAS. Reproduced with permission.}
\label{f:high-energies}
\end{figure}

Also, the already mentioned high-energy missions \agile\ (30~MeV-- 50~GeV), launched on 23 April 2007 and de-orbited in February 2024, and~\fermi\ (20~MeV--300 GeV), launched on 11 June 2008 and still operational, have provided a great contribution to the understanding of the GRB phenomenon at  high energies (see~\cite{Zhang11}). Among~the most relevant results, it certainly merits mentioning the discovery that the onset of the GRB high-energy gamma-ray ($>$100 MeV) emission is delayed with respect to that at lower energies (see left panel of Figure~\ref{f:high-energies}). Another discovered feature is that the gamma-ray spectrum hardens with time from the GRB onset, and a further high-energy spectral component appears in the tail of the gamma-ray light curve (see right panel of Figure~\ref{f:high-energies}). This behavior is confirmed also in the case of very bright events (see~e.g., GRB~130427A \citep{Ackermann14}). All that shows that the high-energy gamma-rays arise from a region and/or a mechanism which is different from that which produces the lower-energy photons. This property is in agreement with the standard model of GRBs, in~which the blast wave that produces the initial, bright prompt emission later collides with the circumburst medium, creating shocks. These external shocks can accelerate the charged particles and produce photons through synchrotron radiation. The~observed high-energy photon energy distribution is consistent with this picture (see e.g.,~\cite{Kouveliotou13}).

%
%

\vspace{-6pt}
 
An important test for establishing whether the afterglow emission is still due to synchrotron at Very High Energies (VHE, $> 100$~GeV)  can be performed with Cerenkov telescopes. To date, only a few GRBs have been detected at these energies: 180720B~\citep{Abdalla19a}, 190114C \citep{Acciari19a}, and 190829A \citep{HESS2021}. From~a multi-wavelength  spectral analysis performed by~\cite{Acciari19b}, it was found that the VHE emission (0.1--1 TeV) detected from GRB~190114C, with the {\em Major Atmospheric Gamma-ray Imaging Cherenkov} (\magic) telescopes~\cite{Acciari19a}, is very likely due to the inverse Compton up-scattering of synchrotron photons. However, the~VHE afterglow emission (0.18--3 TeV) detected from GRB~190829 with the High Energy Stereoscopic System (\hess) is consistent with synchrotron radiation \citep{HESS2021}.

\section{GRB Progenitors in the Post-\sax\ Era}
\label{s:progenitors}
\unskip

\subsection{Long~GRBs}
For long GRBs (T$_{90}> 2$ s), there is still a general consensus that they are mostly due to the core collapse of very massive stars (dubbed as ``collapsars'', see~\cite{Woosley93}). This conclusion is justified from the following facts: (a) the well-established GRB-SN connection; and (b) long GRBs are located in the brightest regions of high SFR galaxies, where the most massive stars are located. However, recently \cite{Rastinejad22,Troja22}, a long GRB (211211A) was associated with a kilonova whose progenitor is a compact binary merger, suggesting that GRBs with long, complex light curves can be spawned also from merger~events. 

\textls[-25]{A different origin for the progenitors of long GRBs was proposed by Ruffini~et~al.~\citep{Ruffini16}~in the case of long XRFs/GRBs, which exhibit two distinct episodes in their light curves. According to these authors, in~these cases, a carbon--oxygen core  undergoes a SN explosion, which triggers a hypercritical accretion onto a NS companion in a tight or more separated binary system. Depending on the system tightness, the~formation or not of a BH is driven, and~a GRB or an XRF event, respectively, is produced.} A~candidate for the formation of a black hole, according to these authors, was GRB~090618 ($E_{iso} = 3 \times 10^{53}$~erg), for which there was an evidence of an SN bump 10 days after the event~\cite{Izzo12}. 

\subsection{Short~GRBs}

Concerning short GRBs, from~the absence of evidence of simultaneous SN explosions  and from the association with galaxies with a wide range of star formation properties (inclusive of low SFR), there is a general consensus that they  are mostly the result of compact binary (e.g., NS-NS, BH-NS) merging as~it was initially supposed. The~association of a gravitational wave signal (GW~170817) with a short GRB (170817) has confirmed this scenario~\cite{Abbott17}. However, also recently, some significant exceptions have been found. One is the case of the short, low-redshift GRB~200826A \citep{Ahumada21,Rossi22} that has been found to be associated with a SN from the optical and NIR bump in the light curve, and~with luminosity and evolution in agreement with several SNe associated with long GRBs. Also, the prompt emission of this event follows the Ep,i–Eiso relation found for long~GRBs. 

From these last discoveries, it is emerging the opportunity to abandon the classification of short and long GRBs and to introduce a new classification: Type I and Type II GRBs. A~useful discussion on this subject can be found in~\cite{Amati21}.

\section{Physics behind GRB Events in the Post-\sax\ Era: Inner~Engine}
\label{engine-physics}

On the basis of the numerous observations of GRBs already made and the study of their properties, the~physics of these events and their afterglows are now at an advanced level of knowledge, with~very comprehensive reviews on these subjects (see~e.g., \cite{Kumar15} and references therein). As~discussed in the cited review, there is a general agreement that GRBs are bright relativistic jets, while the GRB central engine, i.e.,~the mechanism by which this energy is released, is expected to be similar for long and short GRBs, with~two possible main solutions~proposed:
\begin{itemize}
\item
One central engine mechanism assumes that GRBs are powered by mass accretion onto a stellar mass black hole at a very high rate (from a fraction to a few solar masses per second). In~this case, the plasma is extremely hot and forms a thick disk or torus around the central black hole, from~which a GRB jet is launched via three possible mechanisms: (1) a neutrino dominated accretion flow \citep{Chen07}; (2) extraction of  electromagnetic energy by rotation of the black hole via a Poynting flux (mechanism also called Blandford--Znajek process) \citep{Lee00}; and (3) for the case of a highly magnetized accretion disk, the accumulation of energy with the launch of magnetic blobs from the differential accretion of the disk  \citep{Uzdensky06}. 

\item
The other central engine mechanism assumes that a GRB is powered by a rapidly spinning ($ P_0 \sim 1$~ms period), highly magnetized ($B_{surface} \sim 10^{15}$ Gauss) neutron star (``fast magnetar'') when it is spinning down. In~this case, the maximum energy that can be extracted by the magnetar is given by \citep{Kumar15}
\begin{equation*}
E_{rot} \approx \frac{1}{2} I \Omega^2 \approx 2 \times 10^{52} \frac{M}{1.4 M_\odot}(\frac{R}{10^6 \text{cm}})^2 (\frac{P_0}{1~\text{ms}})^{-2}~\text{erg}
\end{equation*}

If the energy released by a GRB is higher than that given by $E_{rot}$, the~fast magnetar mechanism can be ruled out.  
\end{itemize}

It is now well established that most GRB events are the results of the superposition of pulses. Figure~\ref{f:grb010412} is a clear example of this property. Deep studies of the temporal and spectral properties of these pulses and their distribution and evolution  are expected to help to  clarify the engine inner mechanisms that give rise to the GRB jets~\cite{Guidorzi24,Maccary24}. 

A strong contribution to the study of the jet properties of GRBs and their inner engines has been also given by the observations of very strong events, in~particular, from the brightest-of-all-time GRB~221009 recently observed (see~e.g., \cite{O'Connor23,An23}). This event is very unique, with~a huge amount of very-high-energy photons (from 100 GeV to TeV) detected with LHAASO observatory  \citep{Zhang23}. The~combined analysis of these data with those detected at lower energies is expected to be of key importance to obtain information on the dissipation process that gives rise to a GRB \citep{Zhang23}.

\section{GRBs in the Multi-messenger~Era}
\label{multimessenger}

\begin{figure}[H]
	\includegraphics[width=0.8\textwidth]{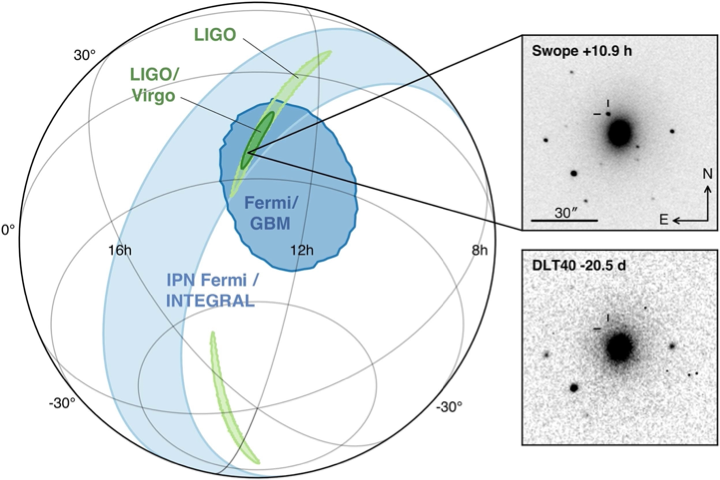}
		\caption[]{{Error} 
 box in the localization of the gravitational event GW~170817 by the LIGO/Virgo detectors and that of the short GRB~170817A by \fermi\ GBM detector. Reprinted from the review by~\cite{Abbott17}.}
	\label{f:170817}
\end{figure}

Until a decade ago, electromagnetic radiation was the only messenger of the phenomena occurring in the universe. Thanks to the enormous technological and financial efforts, in~the last decade, gravitational waves (GWs), high-energy neutrinos ($>$$10$~TeV) and ultra-high-energy cosmic rays ($>$$10^{18}$~eV) have begun to be messengers of astrophysical phenomena, with~the birth of the multi-messenger astrophysics (see historical review by~\cite{Meszaros19}). Evidence of neutrino emission has been detected from blazars \citep{Oikonomou22}, and~(a very suggestive result!) a gravitational wave event (GW~170817) observed with the laser interferometric gravitational wave LIGO/VIRGO detectors has been associated  with a short GRB event (GRB~170817A) \citep{Abbott17}, being almost simultaneously and independently detected. The~localization of these GW and GRB events (see Figure~\ref{f:170817}) was not so constraining, but~their near simultaneity, with~the GRB event occurring only 1.7~s later, was fully consistent with the expectations for the possible production of a GRB event in the case of a NS-NS merging. Indeed, from~the observed properties of the GW event, it was  possible to establish that this event was due to the merging of a binary neutron star system with the NS mass in the range from 0.86 to 2.26 M$_\odot$  at a distance of $40_{-8}^{+8}$~Mpc. Also, the later optical and infrared discovery in the error box region of a kilonova (see~\cite{Pian17}), i.e.,~an outflow rich of high atomic number nuclear elements as a result of the so-called r-process, confirmed this~association. 

After GW~170817, no more GW events have been found to be associated with GRB events, likely due to the fact that almost all of the GW events discovered thus far have been found to be the result of the merging of stellar-mass binary black holes, which are not expected to give rise to electromagnetic radiation. With~the development of more sensitive gravitational detectors, like {\it Einstein Telescope} \citep{Chiummo23}, the~discovery of gravitational signals due to NS-NS merging will be much more likely. 
Binary black hole mergers are also expected in Active Galactic Nuclei \citep{Bartos17}, in~which X-ray and gamma-ray radiation is expected to be~emitted.

%
%
\unskip

\section{The Future for~GRBs}
\label{future}
\unskip

\subsection{Still Open Issues on~GRBs}

Several questions are still open regarding GRB physics and properties. As~above discussed, one still open issue is the central engine (or engines) that powers the GRB events. As~discussed by~\cite{Kumar15}, given their observed non-thermal spectra, GRBs are expected to be efficient cosmic ray accelerators, with~the first-order scenario being the Fermi acceleration mechanism in relativistic shocks (internal and external), alternatively a magnetic reconnection mechanism. The~maximum energy of the accelerated protons can achieve Ultra-High Energies (UHEs), whose interaction can produce neutrinos, in~addition to gamma-ray photons. These neutrinos are being searched with still negative results (see~\cite{Abbasi23}).

An interesting open issue is the determination of the hard X-ray afterglow spectrum. Up~to 10 keV, as we have seen, the~spectrum is well known and can be described by a power-law shape, consistent with synchrotron radiation. The~question is, which is the higher-energy spectrum? an extrapolation of the lower-energy one, and~thus can it still be considered synchrotron radiation? As we have already discussed in Section~\ref{s:spectral-properties}, to date, only for three brightest events, significant measurements of the high-energy spectrum of the GRB afterglow have been obtained. 
The accurate determination of the hard X-ray afterglow spectrum  for a large  sample of GRBs with its extension to higher energies is needed in order to better infer the prevailing emission mechanism at~work.

\subsection{Role of GRBs for Cosmology and Fundamental~Physics}

Given their huge brightness, GRBs can be detected up to the highest redshifts as~already demonstrated with \swift\ GRBs (see Section~\ref{s:swift-results}). This allows to extend the Hubble
diagram substantially beyond the redshift range of Type-Ia SNe and thus to further test the standard cosmology $\Lambda$CDM and its cosmological parameters (see \cite{Moresco22}).

GRBs are also crucial for understanding the formation of the first collapsed structures (Pop-III and early Pop-II stars) and, thus, how and when the first stars formed and how they influence their environment and thus the re-ionization of the intergalactic~medium.

Given their cosmological distance, GRBs also offer the opportunity to settle several questions of fundamental physics.
Among~them are the following:
\begin{itemize}
\item
Test of the Lorentz invariance violation that is expected in some theories of quantum gravity. This test can be performed by measuring the time delay of photons as a function of energy.
\item 
Investigation of the BH physics through signatures in the timing properties of GRB prompt emission.
\item
As also demonstrated by the gravitational wave event GW~170817 associated to  GRB~170817A, these~types of events  provide the unique opportunity to study theories of gravity also beyond general relativity (see~\cite{Jana18}).
\item
Given that, as~discussed above, GRBs are the result of ultra-relativistic shocks with Lorentz factors of several hundreds,  much higher than other possible accelerators like blazars,  GRBs are believed to be strong candidates for particle acceleration to extreme energies that are currently observed among cosmic rays. From~the study of their spectrum, it is possible to constrain the energy distribution of such accelerated particles. It has also been proposed that GRB reverse shocks may serve as potential accelerators of ultra-high-energy cosmic rays \citep{ZhangBT23}.
\item
GRBs are also strong candidates to contribute to the observed UHECRs and high-energy neutrinos because~of the extreme shock acceleration caused by the newborn compact object. An~important role in the production of high-energy neutrinos is expected by low-luminosity GRBs (see~\cite{Meszaros17} and references therein).
\item
Another open issue of fundamental physics is the existence of axion-like particles (ALPs) and their oscillation with photons. Given their cosmological distance, GRBs could provide a tool for testing this issue from the observation of high-energy gamma-ray photons arriving from very high distances. The~recent detection of photons at the energy of several TeV from the GRB221009A ($z = 0.151$) seems to be in contradiction with the expected optical depth for electromagnetic radiation, and~has been recently interpreted  in terms of the existence of ALPs, with~a mass in the range from $10^{-11}$ to $10^{-7}$~eV, that oscillate with photons~\cite{Galanti23}. 
\end{itemize}

GRBs are unique events capable of producing peculiar and well-distinguished electromagnetic radiation up to the highest frequencies that can be detected also from very large~distances.

\section{Future Space Missions and Ground Facilities Devoted to~GRBs}
\label{s:future missions}

\textls[-15]{The still open issues on GRBs and those on fundamental physics that are expected to be solved by GRBs have motivated the development of new space missions and ground~facilities.  }

\subsection{Space~Missions}
\begin{itemize}
    \item 
A mission that has the potentiality of detecting GRBs in the X-ray energy band, where they are less explored, is the Chinese mission  with an international participation {\em Einstein Probe} ({\em EP}), that has been very recently launched (9 January 2024). It has on board two instruments: (1) a Wide-field X-ray Telescope (WXT), based on lobster eye optics, with~a large FOV (60 $\times$ 60 deg) for transient source survey and localization, and~a passband from 0.5 to 4 keV; and (2) a Follow-up X-ray Telescope (FXT), with~\mbox{two units}, each one with Wolter I focusing optics, a~focal length of 1.6~m, a~narrow FOV  (1 $\times$ 1~deg) and an energy passband 0.2--10 keV \citep{Yuan22}. Given the low-energy band of its instruments, the~association of an {\em EP} transient event with a GRB requires the simultaneous detection of the event hard X-ray/soft gamma-ray emission. 
\item
An X-ray/gamma-ray space mission, scheduled to be launched on 24 June 2024, is the Chinese--French mission {\em SVOM} that has on~board a~Gamma-ray Monitor (\mbox{15 keV--5 MeV}), an~X-ray imager and trigger (ECLAIRs, 4--150 keV), a~lobster eye telescope (MXT, 0.2--10~keV) and an optical telescope \citep{Atteia23}.
\item
An already mentioned mission concept for its importance is \theseus\  \citep{Amati21}, that has been approved by ESA for a new phase A study. If~it will be adopted, the~launch is expected to be in 2035. It has on board three instruments, two with a wide FOV (a Soft X-ray Imager (SXI, 0.3--5~keV), based on a lobster eye focusing system, and~ a broad energy band (2~keV--10~MeV) X-ray Gamma-ray Imaging Spectrometer (XGIS) for GRB identification and accurate localization), and~a 70 cm class InfraRed Telescope (IRT, 0.7--1.8 $\upmu$m) with imaging and spectroscopic capabilities (resolving power, $R\approx 400$, through $2'\times 2'$ grism),  for~the GRB  IR counterpart identification and its redshift determination. 
\item 
In the gravitational wave field, a~large mission has been recently approved: the ESA-NASA mission {\em eLISA} (Laser Interferometer Space Antenna), foreseen to be launched in the early 2035s \citep{Amaro-Seoane17}. 
\end{itemize}

\subsection{Ground~Facilities}

Extremely large optical, radio, neutrino, gravitational wave  facilities are already operational or under development, to~extend the band of GRB detection and their multi-messenger~counterparts.  
\begin{itemize}
\item
Among the future optical facilities, I mention the largest ones: the European Extremely Large Telescope {\em EELT} (see~\cite{Ramsay20}), the~US Thirty Meter Telescope {\em TMT} (see~\cite{Skidmore18}, the~American Giant Magellan Telescope ({\em GMT}) (see~\cite{Fanson22}) and the Vera Rubin Observatory \citep{Ivezi19}. In~particular, the~latter will execute a Legacy Survey of Space and Time (LSST) of the entire southern sky every four nights in six different bands during ten~years. 
\item
Also, new radio facilities are under development, particularly the Square Kilometer Array {\em SKA}, that will be the largest radio telescope in the world (see~\cite{MacPherson18}).
\item
In the gravitational wave field, a~gravitational wave European project is being developed (location still not decided): the {\em Einstein Telescope} (see~\cite{Branchesi23}).
\item
In the Very-High-Energy (VHE) gamma-ray field, there are already several operational facilities, like {\em MAGIC} (see~\cite{Ahnen16}) and {\em HESS} (see~\cite{Aharonian23}) already seen, {\em VERITAS} (Very Energetic Radiation Imaging Telescope Array System; see~\cite{Acharyya23}), and~{\em LHAASO} (Large High Altitude Air Shower Observatory; see~\cite{Ma022}). In~addition, a very large project is under development: {\em CTA} (Cerenkov Telescope Array; see~\cite{Hofmann23}).
\item
Also, large neutrino  facilities are already operational, like the underwater telescope {\em ANTARES} (Astronomy with a Neutrino Telescope and Abyss environmental RESearch project; see~\cite{Ageron11}), and~the {\em ICECUBE} neutrino observatory (see~\cite{Aartsen17}). Instead, a next-generation neutrino  facility is the telescope {\em KM3NET} (Cubic Kilometre Neutrino Telescope; see~\cite{Aiello21}).
\end{itemize}

\section{Concluding Soft Gamma-Ray Mission~Concept}
\label{s:astena}

\begin{figure}[H]
	\includegraphics[width=0.6\textwidth]{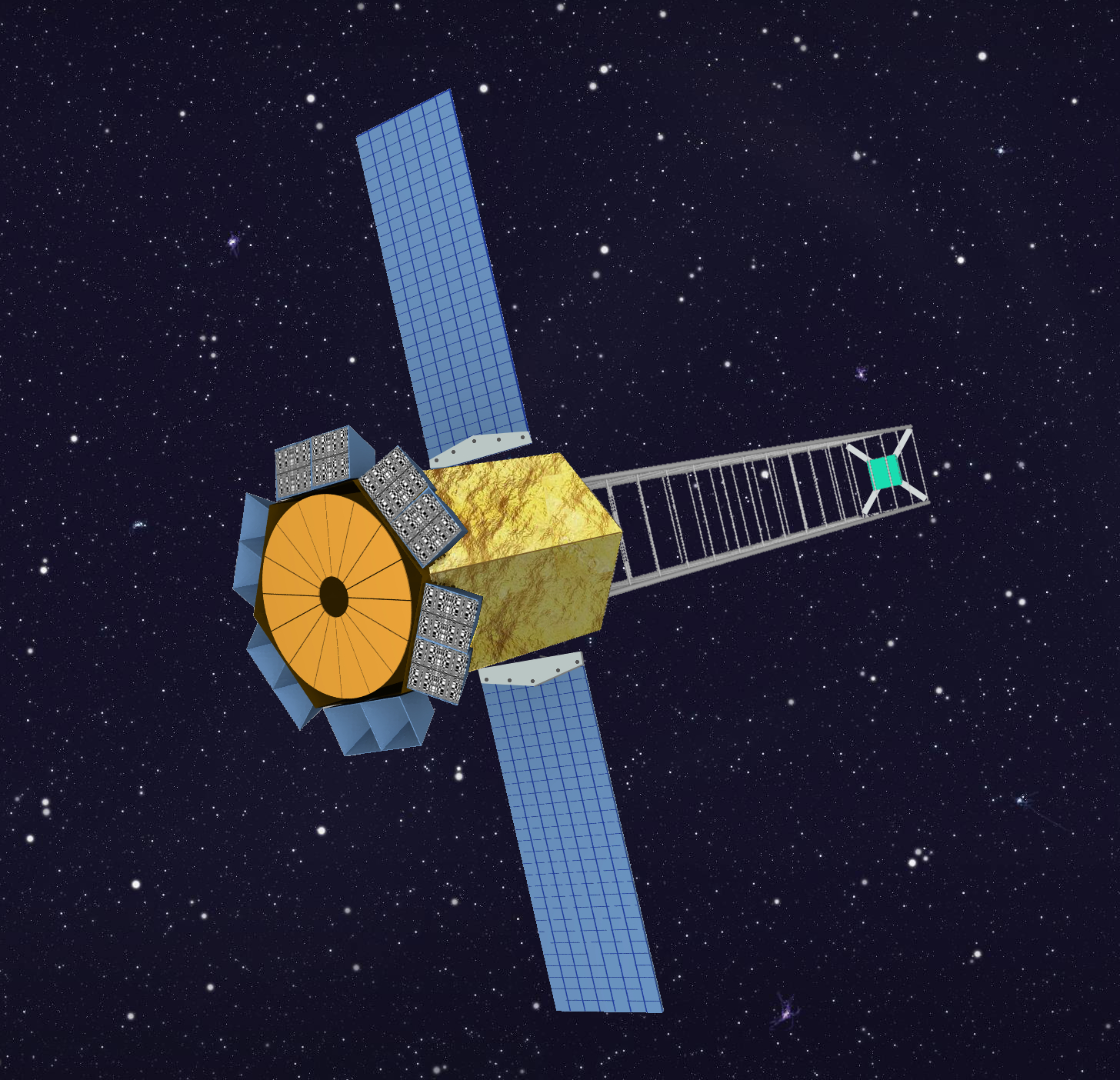}
		\caption[]{{The} 
 {\em ASTENA} mission concept, proposed to ESA for its long-term program ``Voyage 2050''. Reprinted from~\cite{Frontera21}.}
\label{f:astena-in-flight}
\end{figure}

I would like to conclude my review of the future space missions for~deep studies of GRBs by~mentioning  an advanced X-ray/soft gamma-ray mission concept, {\em ASTENA} (Advanced Surveyor of Transient Events and Nuclear Astrophysics) under study by an international collaboration led by the University of Ferrara in the context of the European program {\em AHEAD}. The~foreseen payload (see Figure~\ref{f:astena-in-flight}) includes two main instruments: an array of 12 Wide Field Monitor-Imager Spectrometers (WFM-IS) with a passband from 2~keV to 20~MeV, and~a focusing Narrow Field Telescope (NFT) with a passband from 50 to 700 keV, based on a Laue lens of about 7~m$^2$ collecting area and a 20 m focal length~\cite{Frontera21}. 

Thanks to its large effective area (5800~cm$^2$ at energies $<$30~keV and larger at higher energies), its wide FOV (2 sr) and very good angular resolution (6 arcmin), WFM-IS could perform deep studies of the transient sky, particularly low-luminosity GRBs and other still unknown high-energy transients, contributing so much to future multi-messenger  astronomy~\cite{Guidorzi21}.

On the other side, NFT is expected to improve by orders of magnitude the sensitivity of the best current gamma-ray instruments, inclusive of NuSTAR. NFT will be ideal for studying the still almost unknown soft gamma-ray spectra of GRB afterglows. 

Two white papers, based on the {\em ASTENA} mission concept, were submitted to ESA for its long-term program ``Voyage 2050'' \citep{Frontera21,Guidorzi21}. The~discussed topics were recommended from the Voyage 2050 Senior Committee for a medium size satellite~mission.

All these space and ground facilities are expected to provide further breakthrough results for the GRB astrophysics and, more generally, for~continuing in the advancement of our knowledge of the~Universe.

\vspace{+6pt}

\funding{{This research received no external funding.} 
}

\acknowledgments{{Many} people contributed to the \sax\ discoveries. I wish to thank all of them. I wish to thank the book editors and the three anonymous referees for their very useful suggestions and contributions to improve the paper.  Many thanks to Mauro Orlandini for his help in different aspects of the paper preparation.
Lastly, many thanks to my wife for her patience during my days of work at home for this review, when I should have been available for family needs.}
\conflictsofinterest{{The authors declare no conflict of interest.}} 
 %
%

%

\abbreviations{Abbreviations}{
{The following} 
 abbreviations are used in this manuscript:\\

\noindent 
\begin{tabular}{@{}ll}

ASI     &    Italian Space Agency \\
ASTENA  &   Advanced Surveyor of Transient Events and Nuclear Astrophysics \\
BAT     &    Burst Alert Telescope, aboard \swift \\
BATSE   &    Burst and Transient Source Experiment \\
BH      &    Black Hole \\
CGRO    &    Compton Gamma Ray Observatory \\
FOV     &    Field of View \\
FWHM    &    Full Width at Half Maximum \\
GRB     &    Gamma-Ray Burst \\
HPGSPC  &    High-Pressure Gas Scintillator Proportional Chamber, aboard \sax \\
HRI     &    High-Resolution Imager aboard \rosat \\
IAU     &    International Astronomical Union \\
IBIS    &    Imager on Board Integral Satellite \\
LAD     &    Large Area Detectors of the BATSE experiment \\
LECS    &    Low Energy Concentrator Spectrometer, aboard \sax \\
MECS    &    Medium Energy Concentrator Spectrometer, aboard \sax \\
NASA    &    National Aeronautics and Space Administration \\
NFIs    &    Narrow Field Instruments aboard \sax \\
NIR     &    NearInfraRed band \\
NS      &    Neutron Star \\
PI      &    Principal Investigator \\
PDS     &     Phoswich Detection System, aboard \sax \\
ROSAT   &    ROengten SATellite \\
SAX     &    Satellite Astronomia X (X-ray Astronomy Satellite in Italian) \\
SDC     &    SAX Data Center \\
SFR     &    Star Formation Rate \\
SRON    &    Space Research Of Netherlands \\
THESEUS &    Transient High-Energy Sky and Early Universe Surveyor \\
UVOT    &    Ultraviolet/Optical Telescope, aboard \swift \\
WD      &    White Dwarf \\
WFCs    &     Wide Field Cameras, aboard \sax \\
XRF     &    X-Ray Flash \\
XRR     &    X-Ray Rich  GRB\\
XRT     &    X-Ray Telescope, aboard \swift \\
\end{tabular}
}
%

\begin{adjustwidth}{-\extralength}{0cm}

\reftitle{References}

\PublishersNote{}

\end{adjustwidth}
\end{document}